%% file: Emit-Paper-VB.tex
\documentclass{MICE}
\input GetPackages

\begin{document}
\sloppy
\input{00a-Top-matter}

\makeatletter

\input{MICE-defs}

\parindent 10pt
\pagestyle{plain}
\pagenumbering{arabic}                   
\setcounter{page}{1}

\input{00b-Abstract}

\input{01-Introduction}
\input{02-Emittance-calculation}

\input{03-MICE}
\input{04-MMB}

\input{05-Simulation}

\input{06-BeamSelection}

\input{07-Results}
\input{08-Conclusions}

\input{11-Acknowledgements}

\clearpage
\bibliographystyle{utphys}
\bibliography{Concatenated-bibliography}

\clearpage
\appendix
\section*{The MICE collaboration}
\input{12-AuthorList}

\end{document}

%% file: GetPackages.tex
\usepackage{lineno}
\usepackage{bibentry}
\nobibliography*
\usepackage{times}
\usepackage{bm}         
\usepackage{amsmath,amssymb,bm,slashed} 
\usepackage{amssymb}    
\usepackage{graphicx}   
\usepackage{verbatim}   
\usepackage{color}      
\usepackage{subfigure}  
\usepackage{hyperref}   
\usepackage{enumerate}

\usepackage{fmtcount}   

\usepackage[square,comma,numbers,sort&compress]{natbib}

\usepackage{authblk}	

\usepackage{booktabs}
\usepackage{rotating}
\usepackage[scientific-notation=true]{siunitx}

%% file: 00a-Top-matter.tex
\thispagestyle{empty}

\begin{tabular}{p{0.175\textwidth} p{0.5\textwidth} p{0.225\textwidth}}
  \hspace{-0.8cm}\leftline{\today}                                 &
  \centering{Muon Ionization Cooling Experiment}                   &
  \rightline{RAL-P-2018-005} 
\end{tabular}
\vspace{-1.0cm}\\
\rule{\textwidth}{0.43pt}

\begin{center}
  {\bf
    {\LARGE First particle-by-particle measurement of emittance in the Muon Ionization Cooling Experiment} \\
  }
  \vspace{0.2cm}
  MICE collaboration \\
  \vspace{-0.0cm}
\end{center}

%% file: MICE-defs.tex
\newcommand{\bra}[1]{\ensuremath{\langle #1 |}}   
\newcommand{\ket}[1]{\ensuremath{| #1 \rangle}}   
\newcommand{\bigbra}[1]{\ensuremath{\big\langle #1 \big|}}   
\newcommand{\bigket}[1]{\ensuremath{\big| #1 \big\rangle}}   
\newcommand{\amp}[3]{\ensuremath{\left\langle #1 \,\left|\, #2%
                     \,\right|\, #3 \right\rangle}}  
\newcommand{\sprod}[2]{\ensuremath{\left\langle #1 |%
                     #2 \right\rangle}}  
\newcommand{\ev}[1]{\ensuremath{\left\langle #1 %
                     \right\rangle}} 
\newcommand{\ds}[1]{\ensuremath{\! \frac{d^3#1}{(2\pi)^3 %
                     \sqrt{2 E_\vec{#1}}} \,}} 
\newcommand{\dst}[1]{\ensuremath{\! %
                     \frac{d^4#1}{(2\pi)^4} \,}} 
\newcommand{\tr}{\text{tr}}
\newcommand{\sgn}{\text{sgn}}
\newcommand{\diag}{\text{diag}}
\newcommand{\BR}{\text{BR}}
\newcommand{\gsim}      {\mbox{\raisebox{-0.4ex}{$\;\stackrel{>}{\scriptstyle \sim}\;$}}}
\newcommand{\lsim}      {\mbox{\raisebox{-0.4ex}{$\;\stackrel{<}{\scriptstyle \sim}\;$}}}

\renewcommand{\vec}[1]{{\mathbf{#1}}}
\renewcommand{\Re}{{\text{Re}}}
\renewcommand{\Im}{{\text{Im}}}
\newcommand{\iso}[2]{{\ensuremath{{}^{#2}}\ensuremath{\rm #1}}}
\newcommand{\eps}{{\ensuremath{\epsilon}}}
\newcommand{\draftnote}[1]{{\bf\color{red} \MakeUppercase{#1}}}
\newcommand{\panm}[1]{{\color{blue} #1}}
\providecommand{\abs}[1]{\lvert#1\rvert}
\providecommand{\norm}[1]{\lVert#1\rVert}

\def\parenbar{\mathpalette\p@renb@r}
\def\p@renb@r#1#2{\vbox{%
  \ifx#1\scriptscriptstyle \dimen@.7em\dimen@ii.2em\else
  \ifx#1\scriptstyle \dimen@.8em\dimen@ii.25em\else
  \dimen@1em\dimen@ii.4em\fi\fi \offinterlineskip
  \ialign{\hfill##\hfill\cr
    \vbox{\hrule width\dimen@ii}\cr
    \noalign{\vskip-.3ex}%
    \hbox to\dimen@{$\mathchar300\hfil\mathchar301$}\cr
    \noalign{\vskip-.3ex}%
    $#1#2$\cr}}}

%
\providecommand{\anmne}{\mbox{$\bar\nu_{\mu} \rightarrow \bar\nu_e$}} 
\providecommand{\nmne}{\mbox{$\nu_{\mu}\rightarrow\nu_e$}} 
\providecommand{\anm}{\mbox{$\bar\nu_\mu$}} 
\providecommand{\nm}{\mbox{$\nu_\mu$}}
\providecommand{\nue}{\mbox{$\nu_e$}} 
\providecommand{\ane}{\mbox{$\bar\nu_e$}} 
\providecommand{\enu}{\mbox{$E_\nu$}}
\providecommand{\piz}{\mbox{$\pi^0 $}}
\providecommand{\pip}{\mbox{$\pi^+$}} 
\providecommand{\pim}{\mbox{$\pi^-$}}

%% file: 00b-Abstract.tex
\begin{quotation}
  \noindent
The Muon Ionization Cooling Experiment (MICE) seeks to demonstrate the feasibility of ionization cooling, the technique by which it is proposed to cool the muon beam at a future neutrino factory or muon collider.  The emittance is measured from an ensemble of muons assembled from those that pass through the experiment. A pure muon ensemble is selected using a particle-identification system that can reject efficiently both pions and electrons. The position and momentum of each muon are measured using a high-precision scintillating-fibre tracker in a 4\,T solenoidal magnetic field. This paper presents the techniques used to reconstruct the phase-space distributions in the upstream tracking detector and reports the first particle-by-particle measurement of the emittance of the MICE Muon Beam as a function of muon-beam momentum. 
\end{quotation}

%% file: 01-Introduction.tex
\section{Introduction}
\label{Sect:Intro}

Stored muon beams have been proposed as the source of neutrinos at a
neutrino factory \cite{Geer:1997iz,Apollonio:2002en} and as the means
to deliver multi-TeV lepton-antilepton collisions at a muon collider
\cite{Neuffer:1994bt,Palmer:2014nza}. 
In such facilities the muon beam is produced from the decay of pions
generated by a high-power proton beam striking a target. 
The tertiary muon beam occupies a large volume in phase space. 
To optimise the muon yield for a neutrino factory, and luminosity for
a muon collider, while maintaining a suitably small aperture in the
muon-acceleration system requires that the muon beam be `cooled'
(i.e., its phase-space volume reduced) prior to acceleration.
An alternative approach to the production of low-emittance muon beams
through the capture of $\mu^+\mu^-$ pairs close to threshold in
electron--positron annihilation has recently been
proposed~\cite{Boscolo:2018tlu}.
To realise the luminosity required for a muon collider using this
scheme requires the substantial challenges presented by the
accumulation and acceleration of the intense positron beam, the
high-power muon-production target, and the muon-capture systems to be
addressed.

A muon is short-lived, with a lifetime of 2.2\,$\mu$s in its rest
frame. 
Beam manipulation at low energy ($\leq 1$~GeV) must be
carried out rapidly. 
Four cooling techniques are in use at particle accelerators:
synchrotron-radiation cooling \cite{2012acph.book.....L}; laser
cooling~\cite{PhysRevLett.64.2901,PhysRevLett.67.1238,doi:10.1063/1.329218};
stochastic cooling~\cite{Marriner:2003mn}; and electron
cooling~\cite{1063-7869-43-5-R01}.  
In each case, the time taken to cool the beam is long compared to the
muon lifetime. 
In contrast, ionization cooling is a process that occurs on a short
timescale. A muon beam passes through a material (the absorber), loses
energy, and is then re-accelerated. This cools the beam efficiently with
modest decay losses. 
Ionization cooling is therefore the technique by which it is proposed
to increase the number of particles within the downstream acceptance
for a neutrino factory, and the phase-space density for a muon
collider~\cite{cooling_methods,Neuffer:1983xya,Neuffer:1983jr}. 
This technique has never been demonstrated experimentally and such a
demonstration is essential for the development of future
high-brightness muon accelerators or intense muon facilities.

The international Muon Ionization Cooling Experiment (MICE) has been
designed \cite{MICE-WWW} to perform a full demonstration of transverse
ionization cooling. 
Intensity effects are negligible for most of the cooling channels
conceived for the neutrino factory or muon
collider~\cite{Apollonio:2008aa}.
This allows the MICE experiment to record muon trajectories one
particle at a time. 
The MICE collaboration has constructed two solenoidal spectrometers,
one placed upstream, the other downstream, of the cooling cell.  
An ensemble of muon trajectories is assembled offline, selecting an
initial distribution based on quantities measured in the upstream
particle-identification detectors and upstream spectrometer.  
This paper describes the techniques used to reconstruct the
phase-space distributions in the spectrometers. It presents the first
measurement of the emittance of momentum-selected muon
ensembles in the upstream spectrometer.

%% file: 02-Emittance-calculation.tex
\section{Calculation of emittance}
\label{Sect:EmitCalc}

Emittance is a key parameter in assessing the overall performance of
an accelerator~\cite{Rosenzweig:BeamPhysics}. 
The luminosity achieved by a collider is inversely proportional to the emittance of the colliding beams, and therefore beams with small emittance are required.

A beam travelling through a portion of an accelerator may be described
as an ensemble of particles. 
Consider a beam that propagates in the positive $z$ direction of a
right-handed Cartesian coordinate system, $(x, y, z)$. 
The position of the $i^{\rm th}$ particle in the ensemble is
$\mathbf{r}_i = (x_i, y_i)$ and its transverse momentum is
$\mathbf{p}_{Ti} = (p_{xi}, p_{yi})$; $\mathbf{r}_i$ and
$\mathbf{p}_{Ti}$ define the coordinates of the particle in transverse
phase space.
The normalised transverse emittance, $\varepsilon_N$, of the ensemble
approximates the volume occupied by the particles in four-dimensional
phase space and is given by
\begin{equation}
  \varepsilon_{N} = \frac{1}{m_{\mu}}\sqrt[4]{\det \mathcal{C}} \,,
  \label{Eq:Norm4D}
\end{equation}
where $m_{\mu}$ is the rest mass of the muon, $\mathcal{C}$ is the
four-dimensional covariance matrix,
\begin{equation}
  \mathcal{C} = \left(
    \begin{matrix}
      \sigma_{xx} & \sigma_{xp_x} & \sigma_{xy} & \sigma_{xp_y} \\
      \sigma_{xp_x} & \sigma_{p_{x}p_{x}} & \sigma_{yp_x} & \sigma_{p_{x}p_{y}} \\
      \sigma_{xy}   & \sigma_{yp_x} & \sigma_{yy} & \sigma_{yp_y} \\
      \sigma_{xp_y} & \sigma_{p_{x}p_{y}} & \sigma_{yp_y} & \sigma_{p_yp_y}
    \end{matrix}
  \right) \, ,
  \label{Eq:Var4D}  
\end{equation}
and $\sigma_{\alpha \beta}$, where $\alpha, \beta = x, y, p_{x}, p_{y}$, is given by
\begin{equation}
  \sigma_{\alpha \beta} = \frac{1}{N-1} \left( \Sigma_i^N \alpha_i \beta_i   -
    \frac{\left( \Sigma_i^N \alpha_i \right)\left( \Sigma_i^N \beta_i \right)}{N} \right)
 \, .
  \label{Eq:Cov}
\end{equation}

The MICE experiment was operated such that muons passed through the
experiment one at a time. 
The phase-space coordinates of each muon were measured. 
An ensemble of muons that was representative of the muon beam was
assembled using the measured coordinates. 
The normalised transverse emittance of the ensemble was then calculated
by evaluating the sums necessary to construct the covariance matrix,
$\mathcal{C}$, 
and using equation~\ref{Eq:Norm4D}.

%% file: 03-MICE.tex
\section{The Muon Ionization Cooling Experiment}
\label{Sect:MICE}

The muons for MICE came from the decay of pions produced by an
internal target dipping directly into the circulating proton beam of
the ISIS synchrotron at the Rutherford Appleton Laboratory
(RAL)~\cite{Booth:2012qz, Booth:2016lno}. 
The burst of particles resulting from one target dip is referred to as
a `spill'. 
A transfer line of nine quadrupoles, two dipoles and a superconducting
`decay solenoid' selected a momentum bite and transported the beam into the
experiment~\cite{Bogomilov:2012sr}.
The small fraction of pions that remained in the beam were rejected
during analysis using the time-of-flight hodoscopes, TOF0 and TOF1, and Cherenkov
counters that were installed in the MICE Muon Beam line upstream of the
cooling experiment~\cite{Adams:2013lba, Adams:2015wxp}. 
A `diffuser' was installed at the upstream end of the experiment to
vary the initial emittance of the beam by introducing a changeable amount of tungsten and brass, which are high-$Z$ materials, into
the beam path~\cite{Bogomilov:2012sr}. 

\begin{figure*}[!htbp]
  \begin{center}
    \includegraphics[width=0.95\textwidth]{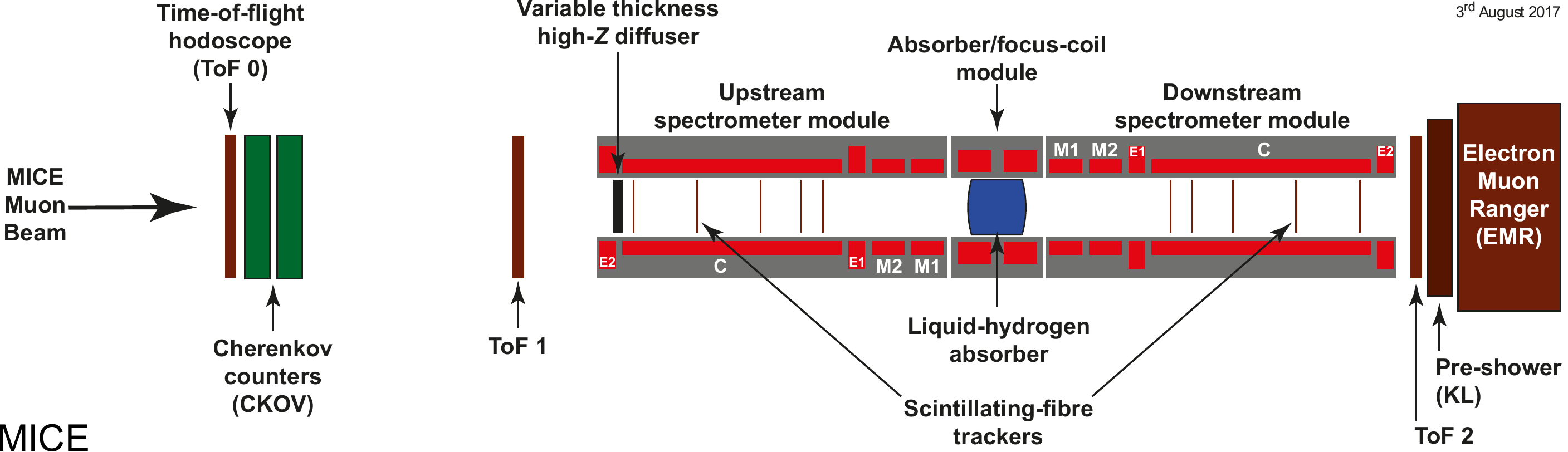}
  \end{center}
  \caption{
    Schematic diagram of the MICE experiment.
    The red rectangles represent the coils of the spectrometer
    solenoids and focus-coil module. 
    The individual coils of the spectrometer solenoids are labelled
    E1, C, E2, M1 and M2.
    The various detectors (time-of-flight
    hodoscopes (TOF0, TOF1)~\cite{Bertoni:2010by,MICE:Note:286:2010}, Cherenkov
    counters~\cite{Cremaldi:2009zj}, scintillating-fibre
    trackers~\cite{Ellis:2010bb}, KLOE-Light (KL)
    calorimeter~\cite{Bogomilov:2012sr,Ambrosino2009239}, and
    Electron Muon Ranger (EMR)~\cite{Asfandiyarov:2016erh,Adams:2015eva}) are also
    represented. The Partial Return Yoke (PRY) is not shown.
  }
  \label{Fig:Overview}
\end{figure*}

A schematic diagram of the experiment is shown in
figure \ref{Fig:Overview}. 
It contained an absorber/focus-coil module sandwiched between two
spectrometer-solenoid modules that provided a uniform magnetic field
for momentum measurement. 
The focus-coil module had two separate windings that were operated
with the same, or opposed, polarities. 
A lithium-hydride or liquid-hydrogen absorber was placed at the
centre of the focus-coil module. An iron Partial Return Yoke (PRY) was
installed around the experiment to contain the field produced by the
solenoidal spectrometers (not shown in figure \ref{Fig:Overview}). 
The PRY was installed at a distance from the beam axis such that its effect on
the trajectories of particles travelling through the experiment was
negligible.

The emittance was measured upstream and downstream of the absorber and
focus-coil module using scintill\-at\-ing-fibre tracking
detectors~\cite{Ellis:2010bb} immersed in the solenoidal field
provided by three superconducting coils E1, C, and E2. 
The trackers were used to reconstruct the trajectories of individual
muons at the entrance and exit of the absorber. 
The trackers were each constructed from five planar stations of
scintillating fibres, each with an active radius of 150\,mm.
The track parameters were reported at the nominal reference plane: the
surface of the scintillating-fibre plane closest to the
absorber~\cite{Dobbs:2016ejn}.
Hall probes were installed on the tracker to measure the
magnetic-field strength in situ.
The instrumentation up- and downstream of the spectrometer modules was
used to select a pure sample of muons. The reconstructed tracks of the selected muons were then used to
measure the muon-beam emittance at the upstream and downstream tracker
reference planes.
The spectrometer-solenoid modules also contained two superconducting
`matching' coils (M1, M2) that were used to match the optics between
the uniform-field region and the neighbouring focus-coil module. The MICE coordinate system is such that the $z$-axis is coincident with the beam direction, the $y$-axis points vertically upward, and the $x$-axis completes a right-handed co-ordinate system.
This paper discusses the measurement of emittance using only the
tracker and beam-line instrumentation upstream of the absorber. The diffuser was fully retracted for the presented data, i.e. no extra material was introduced into the centre of the beam line, so that the incident particle distribution could be assessed. 

%% file: 04-MMB.tex
\section{MICE Muon Beam line}
\label{Sect:Beam}

\begin{figure*}
  \begin{center}
    \includegraphics[width=0.80\textwidth]{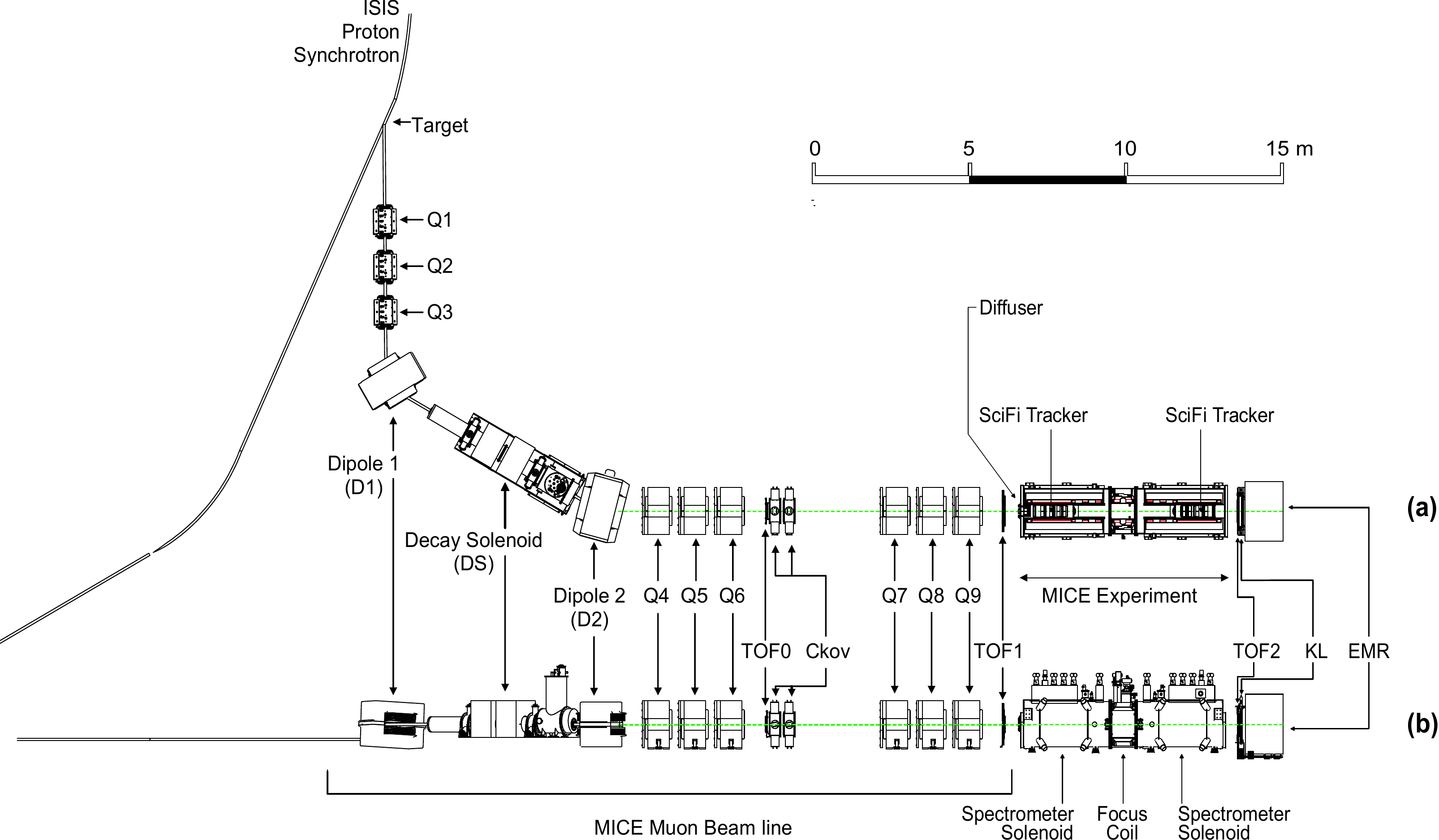}
  \end{center}
  \caption{
    (a) Top and (b) side views of the MICE Muon Beam line, its
    instrumentation, and the experimental configuration. 
    A titanium target dipped into the ISIS proton synchrotron and the
    resultant spill of particles was captured with a quadrupole triplet
    (Q1--3) and transported through momentum-selecting dipoles (D1,
    D2). 
    The quadrupole triplets (Q4--6, Q7--9) transported particles
    to the upstream spectrometer module. 
    The time-of-flight of particles, measured between TOF0 and
    TOF1, was used for particle identification.
  }
  \label{fig:Beamline}
\end{figure*}

The MICE Muon Beam line is shown schematically in
figure~\ref{fig:Beamline}. It was capable of delivering beams with
normalised transverse emittance in the range
$3 \lesssim \varepsilon_N \lesssim 10 $\,mm and mean momentum in the range
$140 \lesssim p_\mu \lesssim 240$\,MeV/$c$ with a root-mean-squared (RMS)
momentum spread of $\sim$20\,MeV/$c$~\cite{Bogomilov:2012sr} after the
diffuser (figure~\ref{Fig:Overview}).

Pions produced by the momentary insertion of a titanium
target \cite{Booth:2012qz,Booth:2016lno} into the ISIS proton beam were
captured using a quadrupole triplet (Q1--3) and transported to a first
dipole magnet (D1), which selected particles of a desired momentum bite
into the 5\,T decay solenoid (DS). 
Muons produced in pion decay in the DS were momentum-selected using a
second dipole magnet (D2) and focused onto the diffuser by a
quadrupole channel (Q4--6 and Q7--9). 
In positive-beam running, a borated polyethylene  absorber of variable
thickness was inserted into the beam just downstream of the decay solenoid 
to suppress the high rate of protons that were produced at the
target~\cite{Blot:2011zz}. 

The composition and momentum spectra of the beams delivered to MICE
were determined by the interplay between the two bending magnets D1 and
D2.
In `muon mode', D2 was set to half the current of D1, selecting
backward-going muons in the pion rest frame. This produced an almost
pure muon beam. 

Data were taken in October 2015 in muon mode at a nominal momentum of
200\,MeV/$c$, with ISIS in operation at 700~MeV. 
These data~\cite{miceDataArchive} are used here to characterise the
properties of the beam accepted by the upstream solenoid with all
diffuser irises withdrawn from the beam.
The upstream E1-C-E2 coils in the spectrometer module were energised
and produced a field of 4\,T, effectively uniform across the tracking
region, while all other coils were unpowered. 
Positively charged particles were selected due to their higher
production rate in 700\,MeV proton-nucleus collisions. 

%% file: 05-Simulation.tex
\section{Simulation}
\label{Sect:Simu}

Monte Carlo simulations were used to determine the accuracy of the
kinematic reconstruction, to evaluate the efficiency of the response
of the scintillating-fibre tracker, and to study systematic
uncertainties. 
A sufficient number of events were generated to ensure that
statistical uncertainties from the simulations were negligible in
comparison to those of the data.

The beam impinging on TOF0 was modelled using
G4beamline~\cite{G4BeamLine:WWW}. 
Particles were produced at the target using a parameterised
particle-production model.
These particles were tracked through the MICE Muon Beam line taking into
account all material in and surrounding the beam line and using
realistic models of the field and apertures of the various magnets. 
The G4beamline simulation was tuned to reproduce the observed particle
distributions at TOF0. 

The MICE Analysis User Software (MAUS)~\cite{MICE:Note:439:2014}
package was used to simulate the passage of particles from TOF0
through the remainder of the MICE Muon Beam line and through the solenoidal lattice.
This simulation includes the response of the instrumentation and used
the input distribution produced using G4beamline. 
MAUS was also used for offline reconstruction and to provide fast real-time
detector reconstruction and data visualisation during MICE running. 
MAUS uses GEANT4~\cite{Ago03,Allison:2006ve} for beam propagation and
the simulation of detector response. ROOT~\cite{Brun:1997pa} was used for data visualisation and for data
storage. 
The particles generated were subjected to the same trigger requirements
as the data and processed by the same reconstruction programs. 

%% file: 06-BeamSelection.tex
\section{Beam selection}
\label{Sect:BeamSelection}

Data were buffered in the front-end electronics and read out at the end
of each spill~\cite{Bogomilov:2012sr}. 
For the reconstructed data presented here, the digitisation of 
analogue signals received from the detectors was triggered by a
coincidence of signals in the PMTs serving a single scintillator
slab in TOF1. 
Any slab in TOF1 could generate a trigger.

\begin{figure*}[!htbp]
  \begin{center}
     \includegraphics[width=0.9\textwidth]{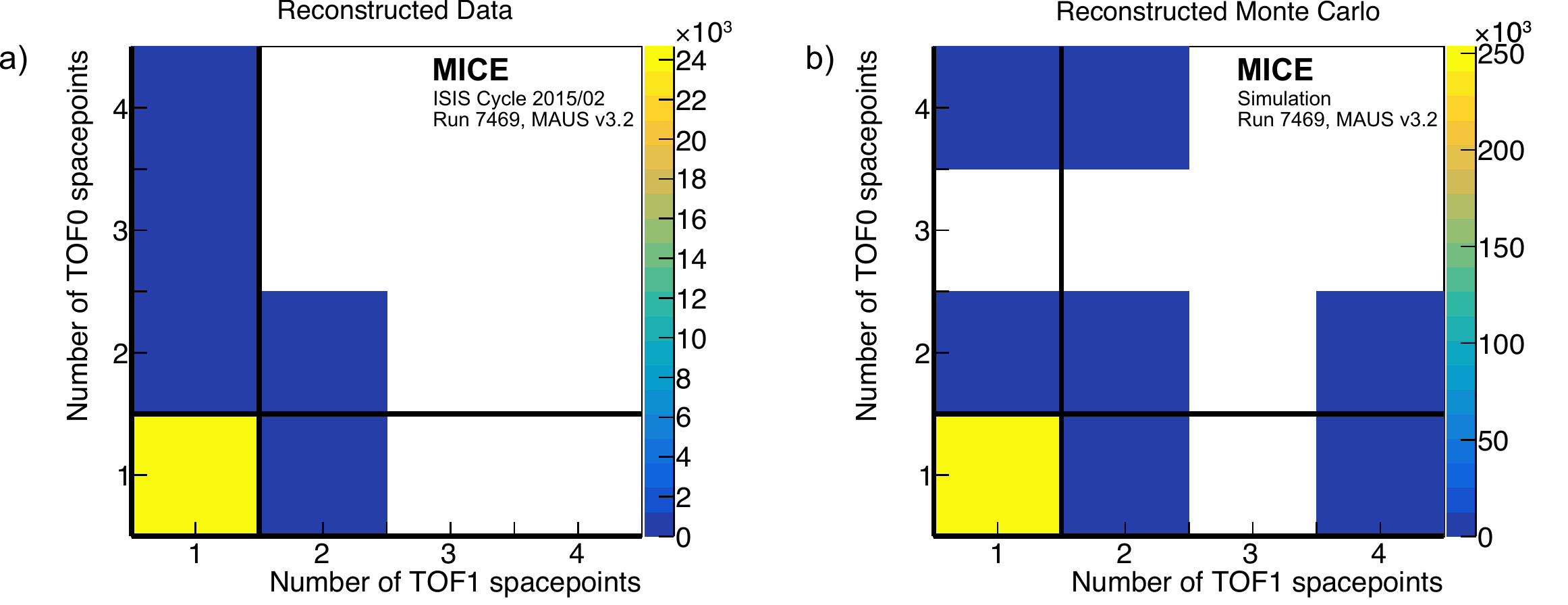} \\
    \includegraphics[width=0.48\textwidth]{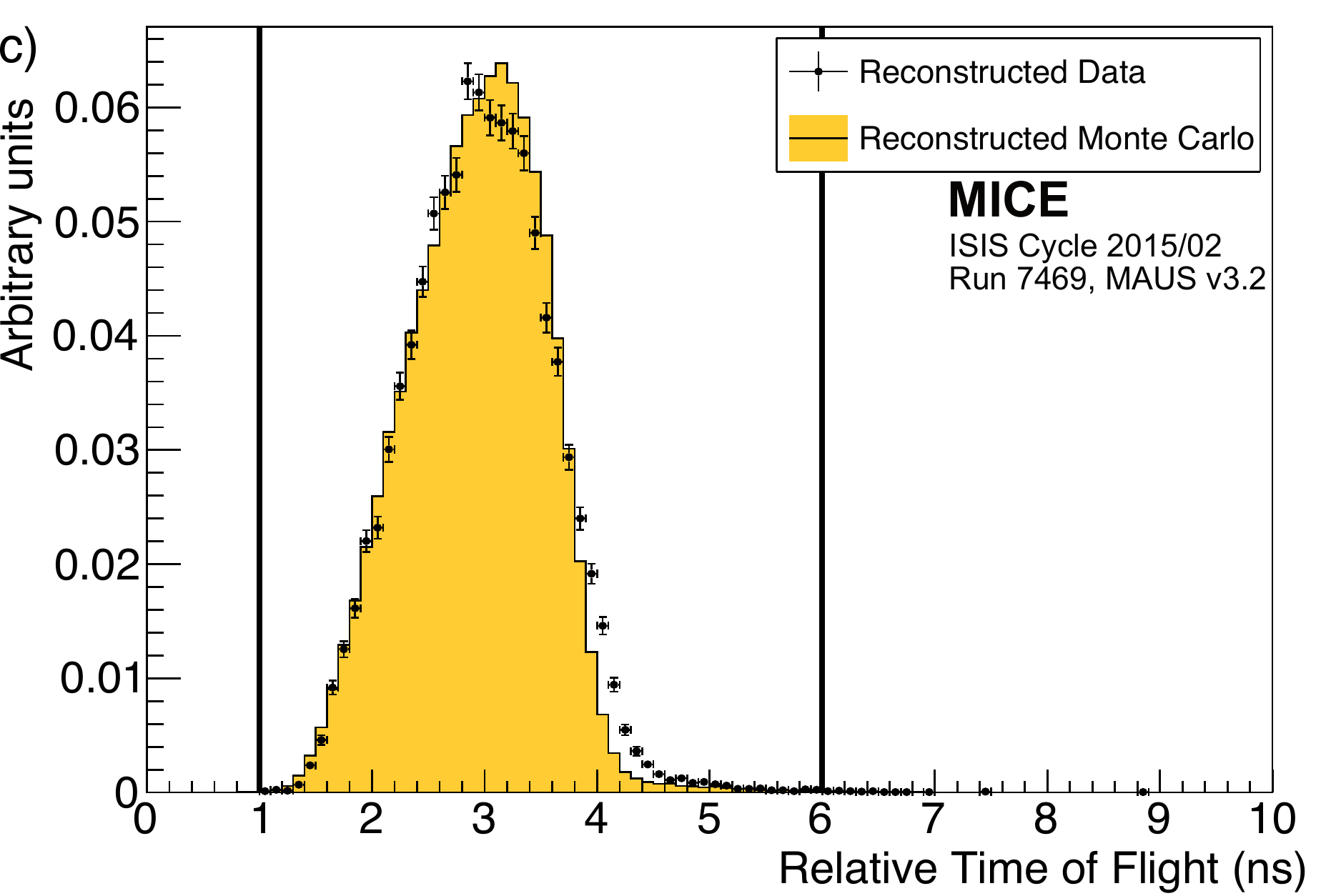} \\
    \includegraphics[width=0.48\textwidth]{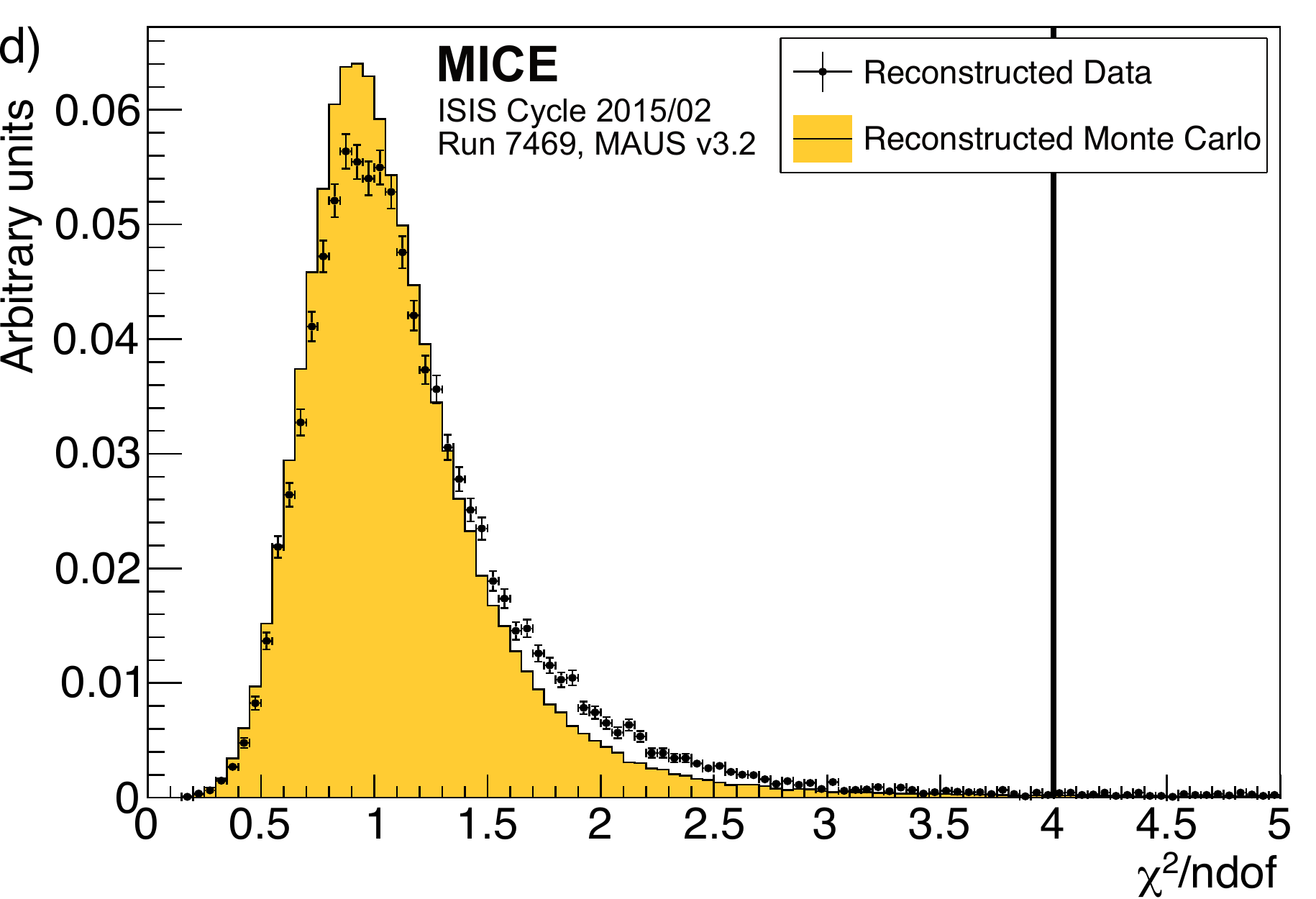} \quad
    \includegraphics[width=0.48\textwidth]{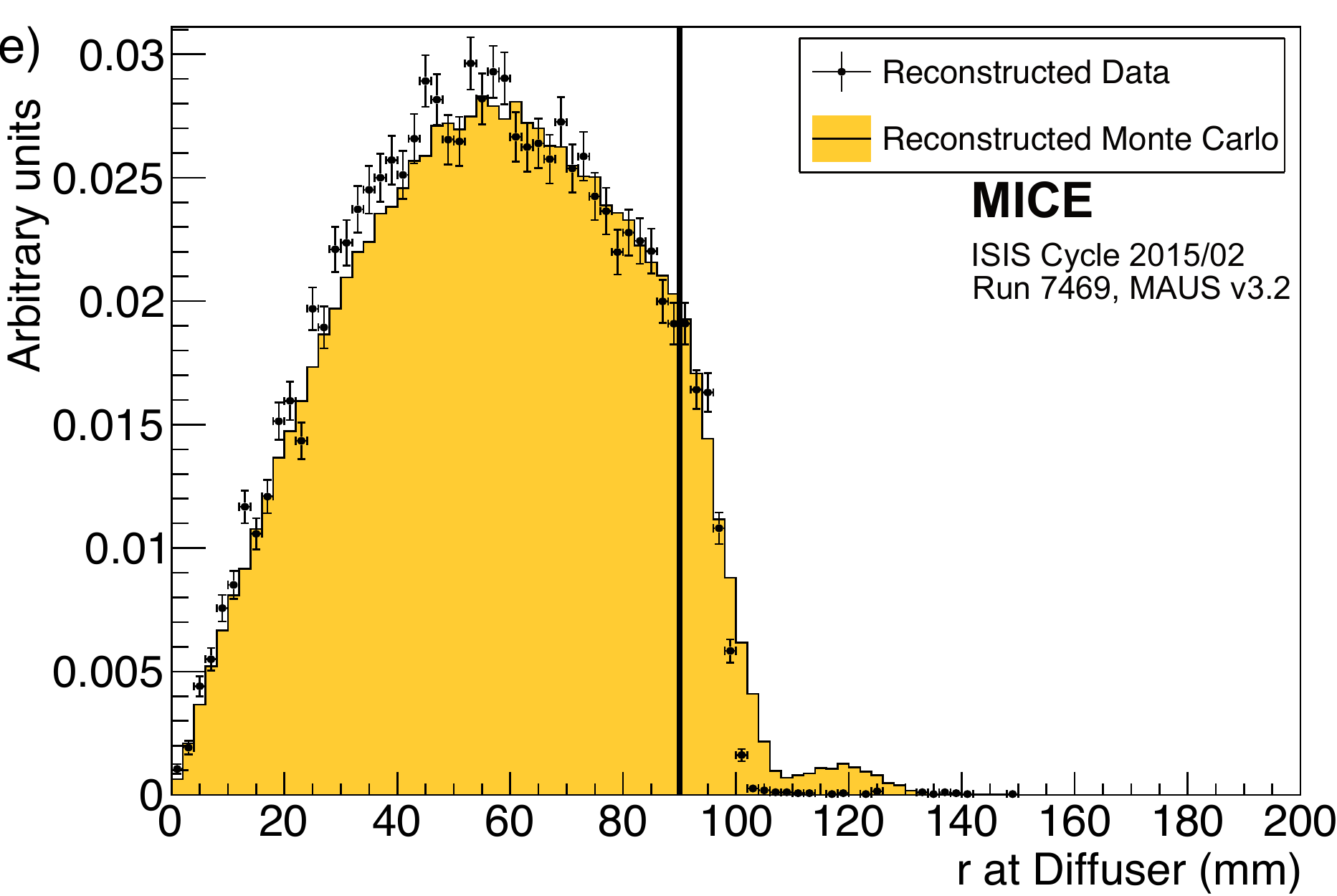}
  \end{center}
  \caption{
    Distribution of the quantities that were used to select the sample used to
    reconstruct the emittance of the beam: 
    a) the number of space-points in TOF0 plotted against the number
    of space-points in TOF1 for reconstructed data, and b)
    reconstructed simulation; c) distribution of the relative
    time-of-flight, $t_{\mathrm{rel}}$; d) distribution of
    $\frac{\chi^2}{N_{\rm DOF}}$; and e) distribution of
    $R_{\rm diff}$. 
    The 1D distributions show reconstructed data as solid (black)
    circles and reconstructed MAUS simulation as the solid (yellow)
    histogram. 
    The solid (black) lines indicate the position of the cuts made on
    these quantities. 
    Events enter these plots if all cuts other than the cut under
    examination are passed. 
  }
  \label{Fig:Cuts}
\end{figure*}

\begin{figure*}[!htbp]
  \begin{center}
    \includegraphics[width=0.90\textwidth]{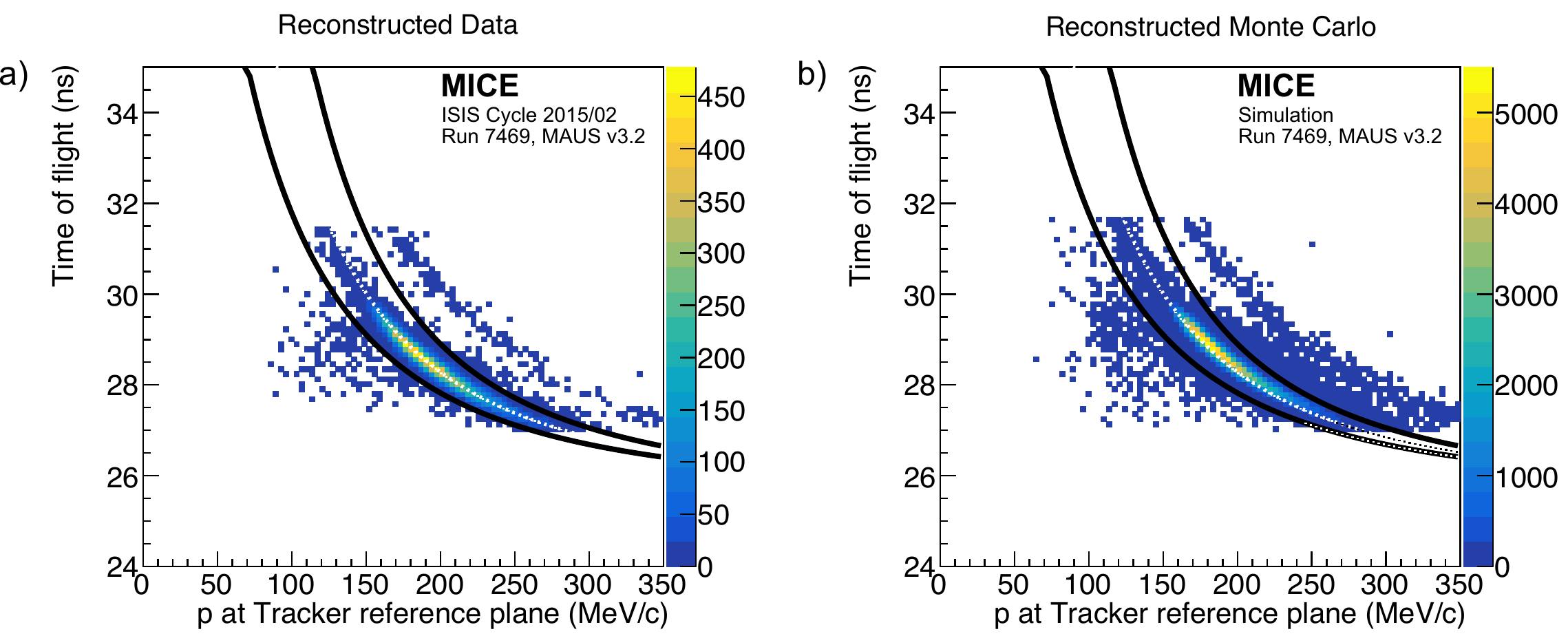}
  \end{center}
  \caption{
    Time of flight between TOF0 and TOF1 ($t_{01}$) plotted as a
    function of the muon momentum, $p$, measured in the upstream
    tracker. 
    All cuts other than the muon hypothesis have been applied. 
    Particles within the black lines are selected. 
    The white dotted line is the trajectory of a muon that loses the
    most probable momentum (20\,MeV/$c$) between TOF1 and the tracker in  
    a) reconstructed data, and b) reconstructed Monte Carlo. 
  }
  \label{Fig:PId}
\end{figure*}

The following cuts were used to select muons passing through the
upstream tracker:
\begin{itemize}
  \item{\it One reconstructed space-point in TOF0 and TOF1}: Each TOF
    hodoscope was composed of two perpendicular planes of scintillator
    slabs arranged to measure the $x$ and $y$ coordinates. 
    A space-point was formed from the intersection of hits in the $x$
    and $y$ projections. 
    Figures~\ref{Fig:Cuts}a and~\ref{Fig:Cuts}b show the hit
    multiplicity in TOF0 plotted against the hit multiplicity in TOF1
    for reconstructed data and reconstructed Monte Carlo respectively.
    The sample is dominated by events with one space-point in both
    TOF0 and TOF1. 
    This cut removes events in which two particles enter the
    experiment within the trigger window.

\item{\it Relative time-of-flight between TOF0 and TOF1, $\mathit{t_{\mathrm{rel}}}$, in the range $\mathit{ 1 \leq t_{\mathrm{rel}} \leq 6}$\,ns}: The time of flight between TOF0 and TOF1, $t_{01}$, was measured relative to
    the mean positron time of flight, $t_e$.
    Figure~\ref{Fig:Cuts}c shows the relative time-of-flight
    distribution in data (black, circles) and simulation (filled histogram). All cuts other than the relative time-of-flight cut have been applied in this figure. The time-of-flight of particles relative to the mean positron time-of-flight is calculated as
    \begin{equation*}
        t_{\mathrm{rel}} = t_{01} - \left( t_{e} + \delta t_{e} \right) \,,
    \end{equation*} 
    where $\delta t_{e}$ accounts for the difference in transit time, or path length travelled, between electrons and muons in the field of the quadrupole triplets~\cite{Adams:2013lba}. This cut removes electrons from the selected ensemble as well as a small number of pions. The data has a longer tail compared to the simulation, which is related to the imperfect simulation of the longitudinal momentum of particles in the beam (see section~\ref{SubSect:Results:PhsSpcPrjctns}).
    
  \item{\it A single track reconstructed in the upstream tracker with
    a track-fit $\mathit{\chi^2}$ satisfying $\mathit{\frac{\chi^2}{N_{\rm DOF}}\leq 4}$}:
    $N_{\rm DOF}$ is the number of degrees of freedom. 
    The distribution of $\frac{\chi^2}{N_{\rm DOF}}$ is shown in
    figure~\ref{Fig:Cuts}d. 
    This cut removes events with poorly reconstructed
    tracks. 
    Multi-track events, in which more than one particle passes
    through the same pixel in TOF0 and TOF1 during the trigger window,
    are rare and are also removed by this cut.
    The distribution of $\frac{\chi^2}{N_{\rm DOF}}$ is broader and peaked
at slightly larger values in the data than in the simulation.
  
  \item{\it Track contained within the fiducial volume of the
    tracker}: The radius of the track measured by the tracker, 
    $R_{\rm track}$, is required to satisfy $R_{\rm track} < 150$\,mm to
    ensure the track does not leave and then re-enter the fiducial 
    volume.
    The track radius is evaluated at 1\,mm intervals between the
    stations.
    If the track radius exceeds 150\,mm at any of these positions, the
    event is rejected. 
 
  \item{\it Extrapolated track radius at the diffuser, $\mathit{R_{\rm diff} \leq 90}$\,mm}:
    Muons that pass through the annulus of the diffuser, which
    includes the retracted irises, lose a substantial amount of
    energy. 
    Such muons may re-enter the tracking volume and be reconstructed
    but have properties that are no longer characteristic of the
    incident muon beam. 
    The aperture radius of the diffuser mechanism (100\,mm) defines the
    transverse acceptance of the beam injected into the experiment. 
    Back-extrapolation of tracks to the exit of the diffuser yields a
    measurement of $R_{\rm diff}$ with a resolution of 
    $\sigma_{R_{\rm diff}} = 1.7$\,mm. 
    Figure~\ref{Fig:Cuts}e shows the distribution of $R_{\rm diff}$, where the difference between data and simulation lies above the accepted radius. These differences are due to approximations in modelling the outer material of the diffuser.
    The cut on $R_{\rm diff}$ accepts particles that passed at least
    $5.9 \sigma_{R_{\rm diff}}$ inside the aperture limit of the diffuser.
  \item{\it Particle consistent with muon hypothesis:}
    Figure \ref{Fig:PId} shows $t_{01}$, the time-of-flight between
    TOF0 and TOF1, plotted as a function of $p$, the momentum
    reconstructed by the upstream tracking detector. 
    Momentum is lost between TOF1 and the reference plane of the tracker in the material of the detectors. A muon that loses the most probable momentum, $\Delta p \simeq 20$\,MeV/$c$, is shown as the dotted (white) line. Particles that are poorly reconstructed, or have passed through support material upstream of the tracker and have lost significant momentum, are excluded by the lower bound. The population of events above the upper bound are ascribed to the passage of pions, or mis-reconstructed muons, and are also removed from the analysis.
\end{itemize}

A total of 24\,660 events pass the cuts listed above. 
Table~\ref{table-cut-summary} shows the number of particles that
survive each individual cut. 
Data distributions are compared to the distributions obtained using
the MAUS simulation in figures~\ref{Fig:Cuts} and~\ref{Fig:PId}. 
Despite minor disagreements, the agreement between the
simulation and data is sufficiently good to give confidence that a
clean sample of muons has been selected.  

The expected pion contamination of the unselected ensemble of particles has been measured to be $\leq 0.4$\,\%\cite{Adams:2015wxp}. Table~\ref{table-cut-mc} shows the number of positrons, muons,
and pions in the MAUS simulation that pass all selection criteria. 
The criteria used to select the muon sample for the analysis presented
here efficiently reject electrons and pions from the Monte Carlo
sample. 

\begin{table*}[!htbp]
  \caption{
    The number of particles that pass each selection criterion. A total of 24\,660 particles pass all of the cuts.
  }
  \label{table-cut-summary}  
  \begin{center}
    \begin{tabular}{@{}lcc@{}}
      \toprule
      Cut                                             & No. surviving particles & Cumulative surviving particles \\ 
      \midrule
      None                                            						& 53\,276   & 53\,276  \\
      One space-point in TOF0 and TOF1                						& 37\,619   & 37\,619  \\
      Relative time of flight in range 1--6\,ns             				& 37\,093   & 36\,658  \\
      Single reconstructed track with $\frac{\chi^2}{N_{\rm DOF}}\leq 4$ 	& 40\,110   & 30\,132  \\
      Track within fiducial volume of tracker         						& 52\,039   & 29\,714  \\
      Extrapolated track radius at diffuser $\leq \,90$\,mm         					& 42\,592   & 25\,310  \\ 
      Muon hypothesis                                 						& 34\,121   & 24\,660  \\
      \midrule
      All                                             						& 24\,660   & 24\,660  \\
      \bottomrule
    \end{tabular}
  \end{center}
\end{table*}

\begin{table*}[!htbp]
  \caption{
    The number of reconstructed electrons, muons, and pions at the
    upstream tracker that survive each cut in the Monte Carlo
    simulation.
    Application of all cuts removes almost all positrons and pions
    in the reconstructed Monte Carlo sample. 
    A total of 253\,504 particles pass all of the described cuts in the
    Monte Carlo simulation.
  }
  \label{table-cut-mc}
  \begin{center}
    \begin{tabular}{@{}lcccc@{}}
      \toprule
      Cut                                             					   & $e$ & $\mu$ & $\pi$ & Total \\ 
      \midrule    
  None                                            					   & 14\,912   & 432\,294 & 1\,610  &  463\,451 \\
  One space-point in TOF0 and TOF1                					   & 11\,222   & 353\,613   & 1\,213  & 376\,528  \\ 
  Relative Time of flight in range 1--6\,ns             			   & 757   & 369\,337   & 1\,217  & 379\,761  \\ 
  Single reconstructed track with $\frac{\chi^2}{N_{\rm DOF}} \leq 4$  & 10\,519   & 407\,276   & 1\,380  & 419\,208  \\
  Track within fiducial volume of tracker         					   &  14\,527  &  412\,857  & 1\,427  & 443\,431  \\ 
  Tracked radius at diffuser $\leq \,90$\,mm      					   &  11\,753  &  311\,076  & 856  &  334\,216 \\ 
  Muon hypothesis (above lower limit)             					   &  3\,225  &  362\,606  & 411  &  367\,340 \\ 
  Muon hypothesis (below upper limit)             					   &  12\,464  &  411\,283  & 379  & 424\,203  \\ 
  Muon hypothesis (overall)                       					   &  2\,724  &  358\,427  &  371 & 361\,576  \\ 
      \midrule
  All                                            					   &  22  &  253\,475  &  5  &  253\,504 \\
      \bottomrule
    \end{tabular}
  \end{center}
\end{table*}

%% file: 07-Results.tex
\section{Results}
\label{Sect:Results}

\subsection{Phase-space projections}
\label{SubSect:Results:PhsSpcPrjctns}

\begin{figure*}[!htbp]
  \begin{center}
    \includegraphics[width=0.9\textwidth]{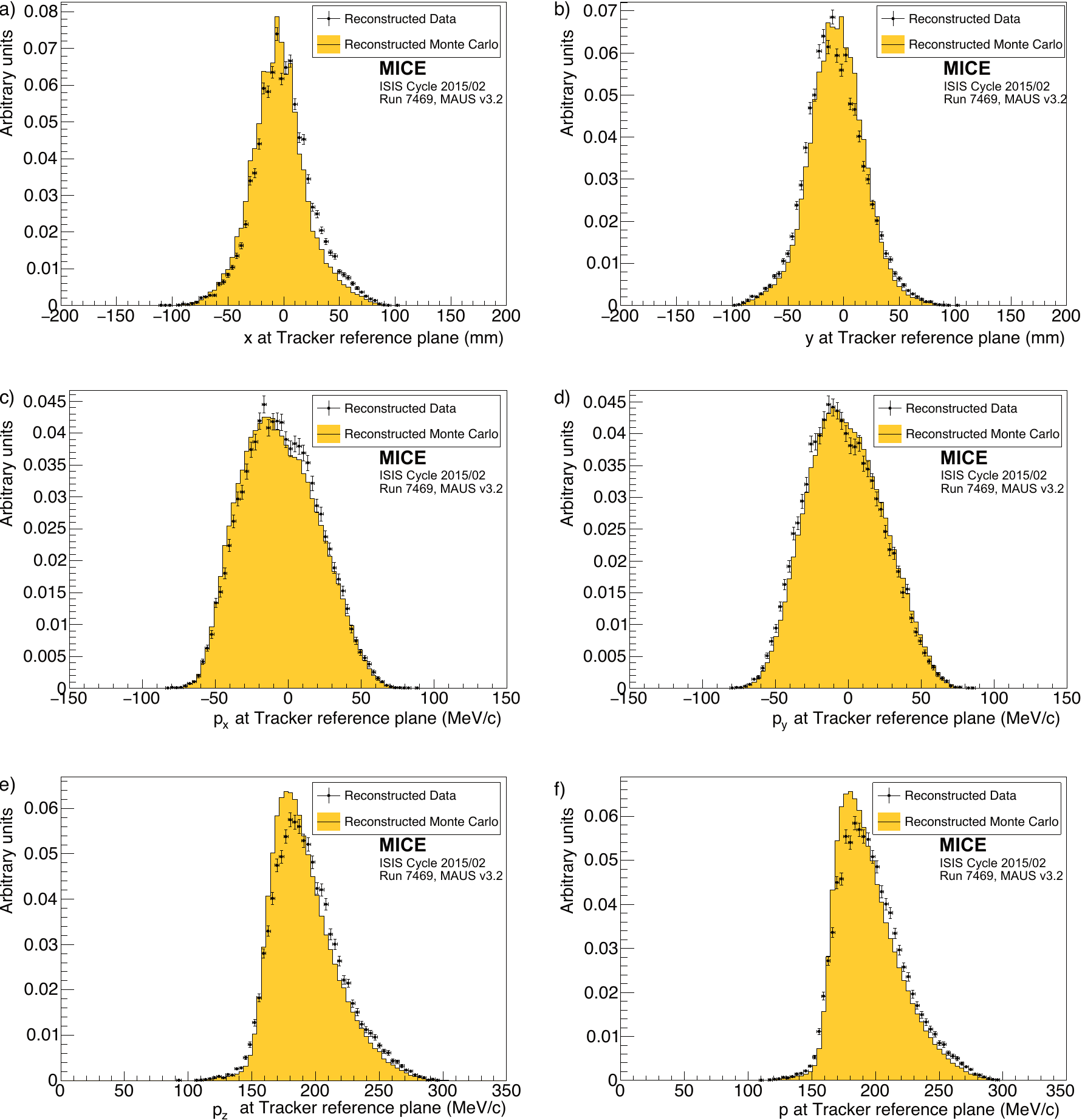}
  \end{center}  
  \caption{
    Position and momentum distributions of muons reconstructed at the reference surface of the upstream tracker:  a) $x$, b) $y$, c) $p_x$, d) $p_y$, e) $p_z$, and f) $p$, the total momentum. The data are shown as the solid circles while the results of the MAUS simulation are shown as the yellow histogram.
  }
  \label{Fig:SngleDstrbtns}
\end{figure*}

The distributions of $x, y, p_x, p_y, p_z$, and $p=\sqrt{p^2_x + p^2_y
+ p^2_z}$ are shown in figure~\ref{Fig:SngleDstrbtns}. 
The total momentum of the muons that make up the beam lie within the
range $140 \lesssim |p| \lesssim 260$\,MeV/$c$. 
The results of the MAUS simulation, which are also shown in
figure~\ref{Fig:SngleDstrbtns}, give a reasonable description of the
data. 
In the case of the longitudinal component of momentum, $p_z$, the data
are peaked to slightly larger values than the simulation. 
The difference is small and is reflected in the distribution of the
total momentum, $p$. 
As the simulation began with particle production from the titanium target, any difference between the simulated and observed particle distributions would be apparent in the measured longitudinal and total momentum distributions. The scale of the observed disagreement is small, and as such the simulation adequately describes the experiment.
The distributions of the components of the transverse phase space ($x,
p_x, y, p_y$) are well described by the simulation. 
Normalised transverse emittance is calculated with respect to the
means of the distributions (equation~\ref{Eq:Var4D}), and so is
unaffected by this discrepancy.

\begin{figure*}[!htbp]
\centering
\includegraphics[width=0.9\textwidth]{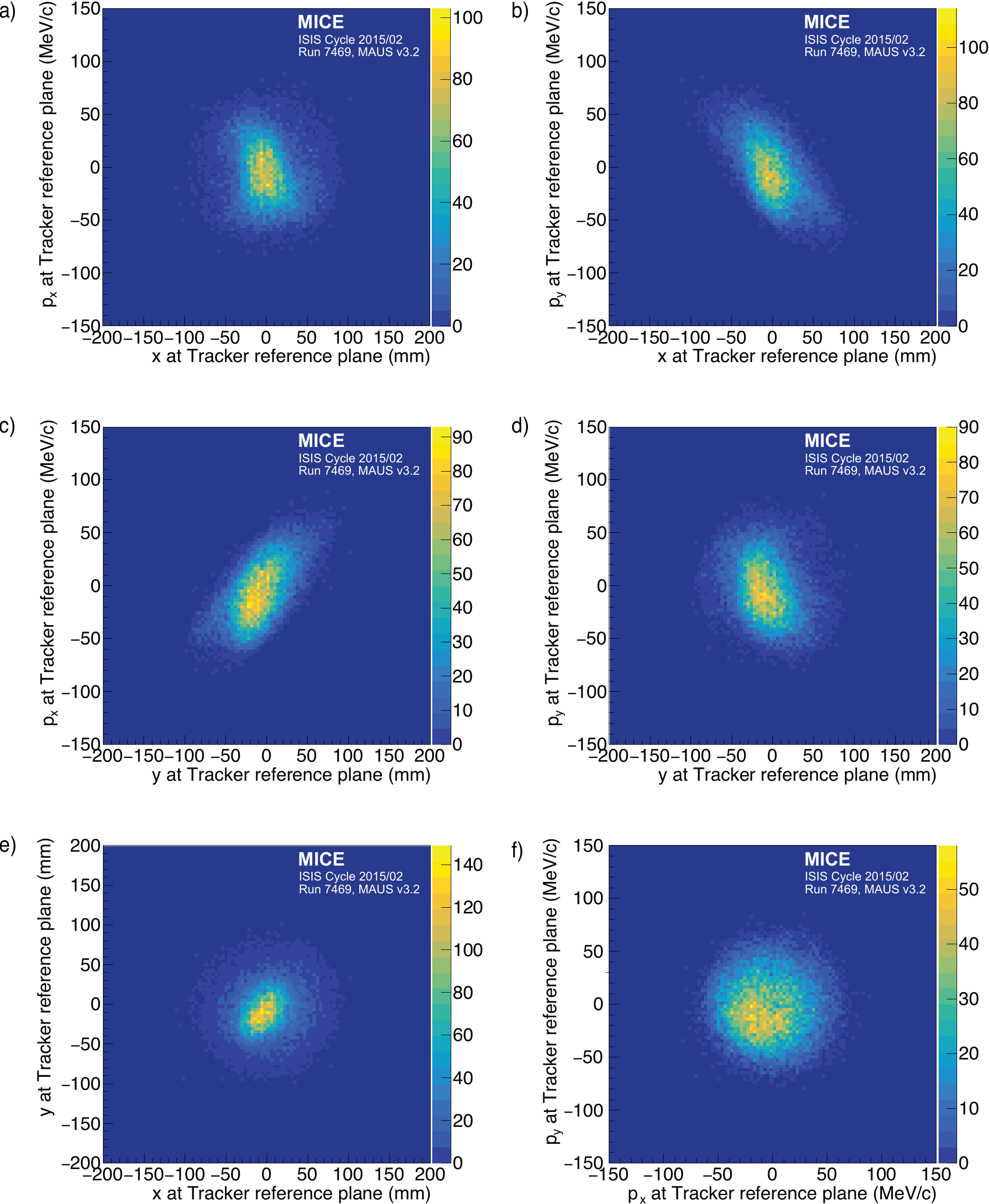}
  \caption{
    Transverse phase space occupied by selected muons transported
    through the MICE Muon Beam line to the reference plane of the
    upstream tracker.
    a) $(x, p_{x})$, b) $(x, p_{y})$.
    c) $(y, p_{x})$, d) $(y, p_{y})$.
    e) $(x, y)$, and f) $(p_{x}, p_{y})$.
  } 
  \label{Fig:Phase-space}
\end{figure*}

The phase space occupied by the selected beam is shown in
figure~\ref{Fig:Phase-space}. 
The distributions are plotted at the reference surface of the upstream
tracker. 
The beam is moderately well centred in the $(x,y)$ plane. 
Correlations are apparent that couple the position and momentum
components in the transverse plane. 
The transverse position and momentum coordinates are also seen to be
correlated with total momentum. The correlation in the $(x, p_{y})$ and $(y, p_{x})$ plane is due to the solenoidal field, and is of the expected order.
The dispersion and chromaticity of the beam are discussed further in
section~\ref{SubSect:Results:BnnngNDsprsn}.

\subsection{Effect of dispersion, chromaticity, and binning in longitudinal momentum}
\label{SubSect:Results:BnnngNDsprsn}

\begin{figure}[!htbp]
  \centering
  \includegraphics[height=0.8\textwidth]{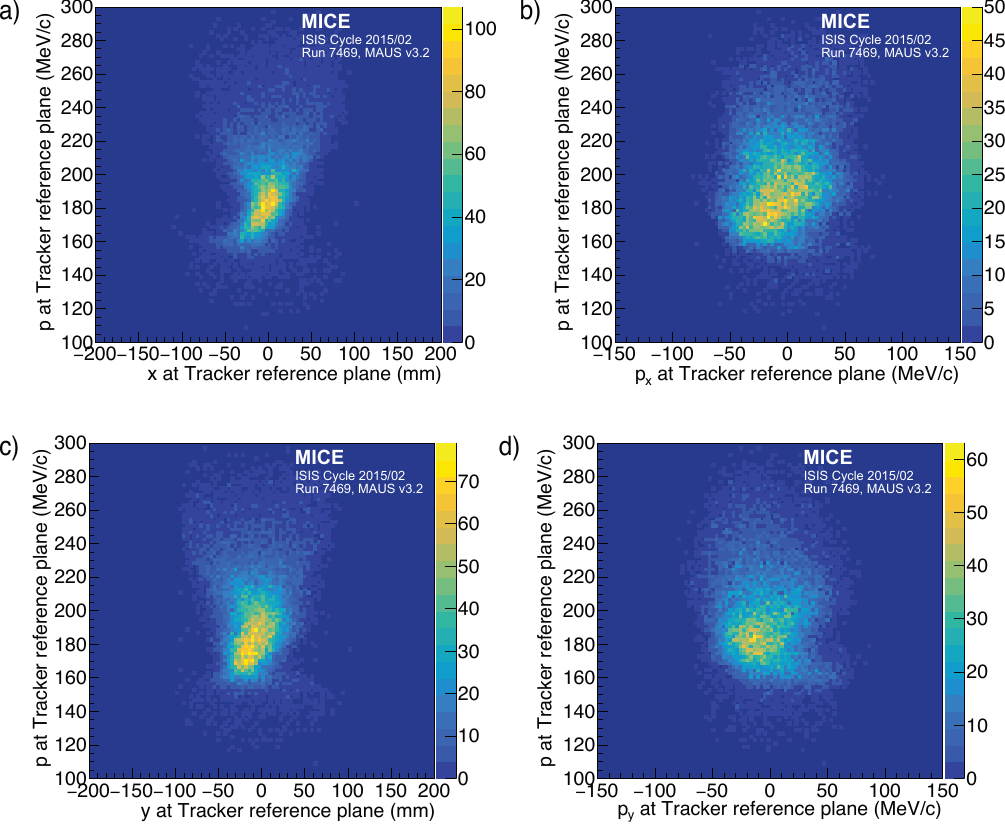}
  \caption{
    The effect of dispersion, the dependence of the components of transverse phase space on the momentum, $p$, is shown at the reference surface of the upstream tracker: a) $(x, p)$; b) $(p_{x},p)$; c) $(y, p)$; d) $(p_{y}, p)$.
  } 
  \label{Fig:Phase-space-dispersion}
\end{figure}

Momentum selection at D2 introduces a correlation, dispersion, between the position and
momentum of particles. 
Figure~\ref{Fig:Phase-space-dispersion} shows the transverse position
and momentum with respect to the total momentum, $p$, as measured at the
upstream-tracker reference plane. 
Correlations exist between all four transverse phase-space
co-ordinates and the total momentum. 

Emittance is calculated in $10$\,MeV/$c$ bins of total momentum in the range $185 \leq p \leq 255$\,MeV/$c$. This bin size was chosen as it is commensurate with the detector resolution. 
Calculating the emittance in momentum increments makes the effect of the optical mismatch, or chromaticity, small compared to the statistical uncertainty. The range of $185 \leq p \leq 255$\,MeV/$c$ was chosen to maximise the number of particles in each bin that are not scraped by the aperture of the diffuser.

\subsection{Uncertainties on Emittance Measurement}\label{Sec:Uncertainty}

\subsubsection{Statistical uncertainties}

The statistical uncertainty on the emittance in each momentum bin is
calculated as
$\sigma_{\varepsilon}=\frac{\varepsilon}{\sqrt{2N}}$~\cite{MICE:Note:268:2009,MICE:Note:341:2011,Unpublished:MICE:Note:2015},
where $\varepsilon$ is the emittance of the ensemble of muons in the
specified momentum range and $N$ is the number of muons in that
ensemble. 
The number of events per bin varies from $\sim 4\,000$ for $p \sim 190$\,MeV/$c$ to $\sim 700$ for $p \sim 250 $\,MeV/$c$.

\subsection{Systematic uncertainties}
\label{SubSect:Results:SysUncert}

\subsubsection{Uncorrelated systematic uncertainties}
\label{SubSubSect:Results:SysUncert:Uncorr}

Systematic uncertainties related to the beam selection were estimated
by varying the cut values by an amount corresponding to the RMS
resolution of the quantity in question. 
The emittance of the ensembles selected with the changed cut values
were calculated and compared to the emittance calculated using the
nominal cut values and the difference taken as the uncertainty due to
changing the cut boundaries. 
The overall uncertainty due to beam selection is summarised in
table~\ref{table-error-summary}. 
The dominant beam-selection uncertainty is in the selection of
particles that successfully pass within the inner 90\,mm of the
diffuser aperture. 

Systematic uncertainties related to possible biases in calibration
constants were evaluated by varying each calibration constant by its
resolution. 
Systematic uncertainties related to the reconstruction algorithms were
evaluated using the MAUS simulation. 
The positive and negative deviations from the nominal emittance were
added in quadrature separately to obtain the total positive and
negative systematic uncertainty. 
Sources of correlated uncertainties are discussed
below.

\subsubsection{Correlated systematic uncertainties}
\label{SubSubSect:Results:SysUncert:Corr}

\begin{figure}[!ht]
  \begin{center}
    \includegraphics[width=0.9\columnwidth]{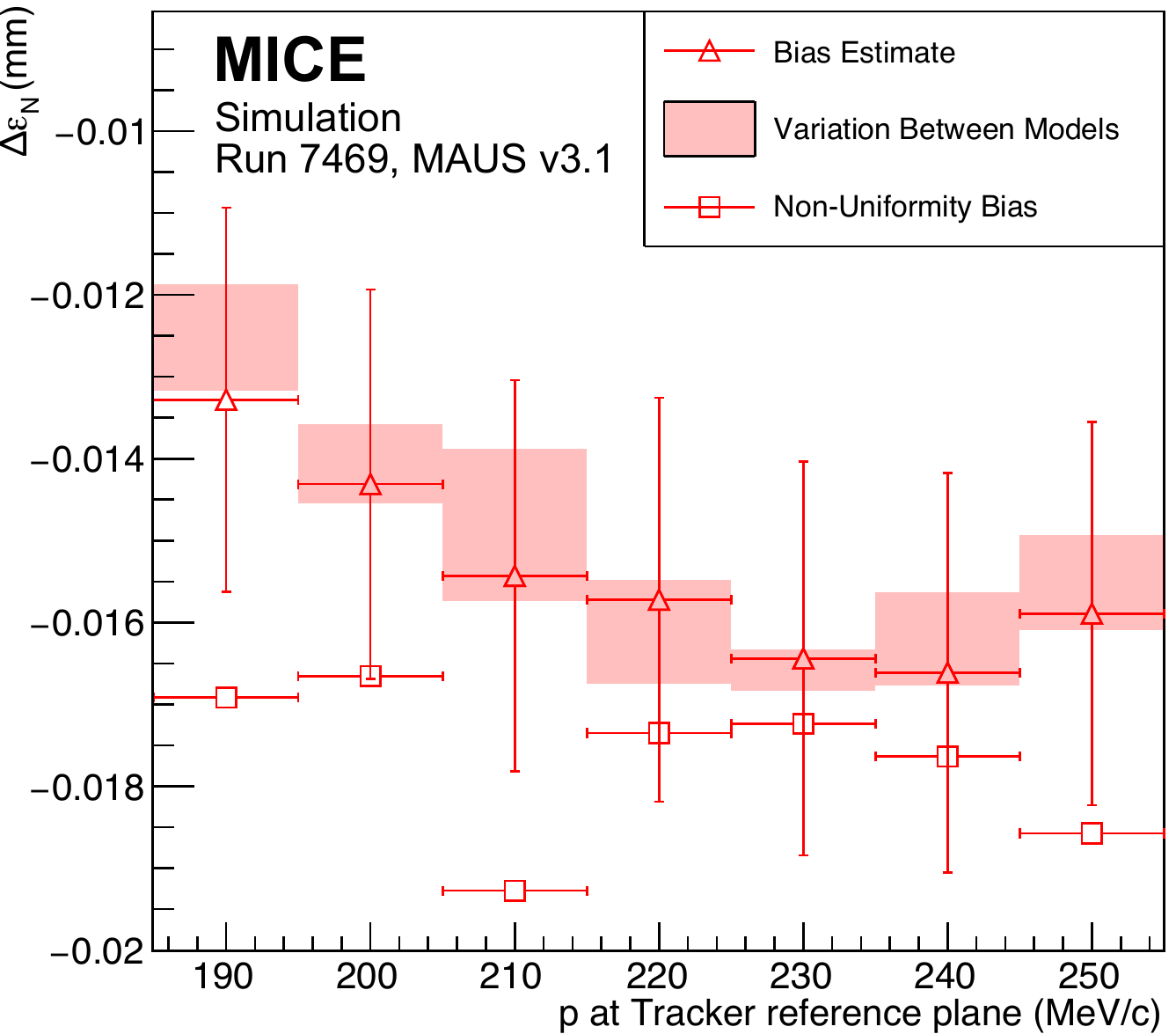}
  \end{center}
  \caption{
    The systematic bias and uncertainty on the reconstructed emittance
    under different magnetic field model assumptions. 
    The bias estimate (open triangles) includes the non-uniformity
    bias (open squares).
    The variation between the models (see text) is indicated by the
    shaded bands.
  }
  \label{bias-fig}
\end{figure}

Some systematic uncertainties are correlated with the total momentum,
$p$. 
For example, the measured value of $p$ dictates the momentum bin to
which a muon is assigned for the emittance calculation. 
The uncertainty on the emittance reconstructed in each bin has been
evaluated by allowing the momentum of each muon to fluctuate around
its measured value according to a Gaussian distribution of width equal
to the measurement uncertainty on $p$. 
In table~\ref{table-error-summary} this uncertainty is listed as
`Binning in $p$'. 

A second uncertainty that is correlated with total momentum is the
uncertainty on the reconstructed $x, p_x, y$, and $p_y$. 
The effect on the emittance was evaluated with the same procedure
used to evaluate the uncertainty due to binning in total momentum. 
This is listed as `Tracker resolution' in
table~\ref{table-error-summary}. 

Systematic uncertainties correlated with $p$ are primarily due to the
differences between the model of the apparatus used in the
reconstruction and the hardware actually used in the experiment.
The most significant contribution arises from the magnetic field
within the tracking volume. 
Particle tracks are reconstructed assuming a uniform solenoidal field,
with no fringe-field effects. 
Small non-uniformities in the magnetic field in the tracking volume
will result in a disagreement between the true parameters and the
reconstructed values. 
To quantify this effect, six field models (one optimal and five
additional models) were used to estimate the deviation in
reconstructed emittance from the true value under realistic
conditions. 
Three families of field model were investigated, corresponding to the
three key field descriptors: field scale, field alignment, and field
uniformity. 
The values of these descriptors that best describe the Hall-probe
measurements were used to define the optimal model and the uncertainty
in the descriptor values were used to determine the $1\sigma$
variations.

\subsubsection{Field scale}
Hall-probes located on the tracker provided measurements of the
magnetic field strength within the tracking volume at known
positions. 
An optimal field model was produced with a scale factor of 0.49\,\% that
reproduced the Hall-probe measurements.
Two additional field models were produced which used scale factors
that were one standard deviation, $\pm0.03$\,\%, above and below the nominal value.

\subsubsection{Field alignment}
A field-alignment algorithm was developed based on the determination of the orientation of the field with respect to the mechanical axis of the tracker using coaxial tracks with $p_{T} \approx 0$~\cite{Hunt-note-in-progress}. The field was rotated with respect to the tracker by $1.4\pm0.1$\,mrad about the $x$-axis and $0.3\pm0.1$\,mrad about the $y$-axis.
The optimal field model was created such that the simulated alignment is in agreement with the measurements. Two additional models that vary the alignment by one standard deviation were also produced.

\subsubsection{Field uniformity}
A COMSOL \cite{comsol} model of the field was used to generate the optimal model which includes the field generated by each coil using the `as-built' parameters and the partial return yoke. A simple field model was created using only the individual coil geometries to provide additional information on the effect of field uniformity on the reconstruction.
The values for the simple field model were normalised to the Hall-probe measurements as for the other field models. This represents a significant deviation from the COMSOL model, 
but demonstrates the stability of the reconstruction with respect to changes in field uniformity, as the variation in emittance between all field models is small (less than 0.002\,mm).

For each of the 5 field models, multiple 2000-muon ensembles were
generated for each momentum bin. 
The deviation of the calculated emittance from the true emittance was
found for each ensemble. 
The distribution of the difference between the ensemble emittance and
the true emittance was assumed to be Gaussian with mean $\varepsilon$
and variance $s^2=\sigma^2+\theta^2$, where $\sigma$ is the
 statistical uncertainty and $\theta$ is an additional
systematic uncertainty. 
The systematic bias for each momentum bin was then calculated
as~\cite{Lyons:0305-4470-25-7-035}
\begin{equation}
  \Delta \varepsilon_{N} = \left< \varepsilon \right> - \varepsilon_{\textrm{true}}\,, 
\end{equation}
where $\varepsilon_{\textrm{true}}$ is the true beam emittance in that
momentum bin and $\left< \varepsilon \right>$ is the mean emittance
from the $N$ ensembles. 
The systematic uncertainty was calculated assuming that the distribution of residuals of $\varepsilon_i$ from the mean, $\left<\varepsilon\right>$, satisfies a $\chi^2$ distribution with $N-1$ degrees of freedom,
\begin{equation}
  \chi^2_{N-1} = \sum_i^N \frac{(\varepsilon_i - \left<\varepsilon\right> )^2}{\sigma^2 + \theta^2}\, ,
\end{equation}
and $\theta$ was estimated by minimising the expression 
$(\chi^2_{N-1} - (N-1))^2$~\cite{Lyons:0305-4470-25-7-035}.

The uncertainty, $\theta$, was consistent with zero in all momentum
bins, whereas the bias, $\Delta \varepsilon_{N}$, was found to be momentum dependent as
shown in figure~\ref{bias-fig}. 
The bias was estimated from the mean difference between the
reconstructed and true emittance  values using the optimal field
model. 
The variation in the bias was calculated from the range of values
reconstructed for each of the additional field models. 
The model representing the effects of non-uniformities in the field
was considered separately due to the significance of the deviation
from the optimal model. 

The results show a consistent systematic bias in the reconstructed emittance of $\approx -0.015$\,mm that is a function of momentum (see table~\ref{table-error-summary}). The absolute variation in the mean values between the models that were used was smaller than the expected statistical fluctuations, demonstrating the stability of the reconstruction across the expected variations in field alignment and scale. The effect of the non-uniformity model was larger but still demonstrates consistent reconstruction. The biases calculated from the optimal field model were used to correct the emittance values in the final calculation (Section~\ref{SubSect:Results:Emmttnc}).


\setlength\tabcolsep{3pt}

\begin{sidewaystable*}[!htbp]
  \centering
  \caption{
    Emittance together with the statistical and systematic uncertainties and biases as a function of mean total momentum, 
    $\langle p \rangle$.
  }
\label{table-error-summary}
\begin{tabular}{llcccccccc@{}}
\toprule
Source                                             & \multicolumn{7}{c}{$\langle p \rangle$ (MeV/$c$)} \\ 
  &   190 &   200 &   210 &   220 &   230 &   240 & 250 \\
\midrule
Measured emittance (mm\,rad) 
                &   3.40
                &   3.65
                &   3.69
                &   3.65 
                &   3.69
                &   3.62 
                &   3.31  \\
\midrule
\midrule
Statistical uncertainty
                        &   $\pm$\num{3.8e-2}  
                        &   $\pm$\num{4.4e-2} 
                        &   $\pm$\num{5.0e-2} 
                        &   $\pm$\num{5.8e-2} 
                        &   $\pm$\num{7.0e-2} 
                        & $\pm$\num{8.4e-2}   
                        & $\pm$\num{9.2e-2}   
                        \\

\midrule
\midrule

Beam selection:    
                  &  
                  &  
                  &  
                  &  
                  &  
                  & 
                  & 
                  \\
                  
~~~~~Diffuser aperture
                  &  $\substack{\num{+4.9e-2}\\-\num{3.5e-2}}$ 
                  &  $\substack{\num{+5.3e-2}\\\num{-5.1e-2}}$ 
                  &  $\substack{\num{+4.9e-2}\\\num{-5.7e-2}}$ 
                  &  $\substack{\num{+4.7e-2}\\\num{-5.0e-2}}$                
                  &  $\substack{\num{+4.2e-2}\\\num{-3.5e-2}}$ 
                  & $\substack{\num{+11.0e-2}\\\num{-5.0e-2}}$                   
                  & $\substack{\num{+4.4e-2}\\\num{-9.6e-2}}$    
                  \\
                  
~~~~~$\frac{\chi^2}{N_{\rm DOF}}\leq 4$
                  &  $\substack{\num{+5.1e-3}\\\num{-4.8e-3}}$ 
                  &  $\substack{\num{+2.0e-3}\\\num{-1.3e-3}}$ 
                  &  $\substack{\num{+1.0e-2}\\\num{-1.8e-3}}$ 
                  &  $\substack{\num{+4.1e-3}\\\num{-3.3e-3}}$                
                  &  $\substack{\num{+1.2e-3}\\\num{-2.8e-4}}$ 
                  & $\substack{\num{+5.5e-3}\\\num{-6.5e-3}}$                   
                  & $\substack{\num{+7.9e-3}\\\num{-4.7e-4}}$    
                  \\

~~~~~Muon hypothesis
                  &  $\substack{\num{+4.5e-3}\\\num{-3.2e-3}}$ 
                  &  $\substack{\num{+2.2e-4}\\\num{-6.8e-3}}$ 
                  &  $\substack{\num{+6.4e-3}\\\num{-8.8e-4}}$ 
                  &  $\substack{\num{+3.1e-3}\\\num{-4.7e-3}}$                
                  &  $\substack{\num{+1.4e-3}\\\num{-1.1e-2}}$ 
                  & $\substack{\num{+2.6e-3}\\\num{-6.7e-2}}$                   
                  & $\substack{\num{+1.3e-3}\\\num{-4.1e-3}}$    
                  \\

~~~~~Beam selection (Overall)    
                  &  $\substack{\num{+4.9e-2}\\\num{-3.6e-2}}$ 
                  &  $\substack{\num{+5.3e-2}\\\num{-5.2e-2}}$ 
                  &  $\substack{\num{+5.0e-2}\\\num{-5.8e-2}}$ 
                  &  $\substack{\num{+4.7e-2}\\\num{-5.0e-2}}$                
                  &  $\substack{\num{+4.2e-2}\\\num{-3.9e-2}}$ 
                  & $\substack{\num{+1.1e-1}\\\num{-8.4e-2}}$                   
                  & $\substack{\num{+4.5e-2}\\\num{-9.6e-2}}$    
                  \\

\midrule
Binning in $p$ 
                &   $\pm\num{1.8e-2}$  
                &   $\pm\num{2.1e-2}$  
                &   $\pm\num{2.3e-2}$  
                &   $\pm\num{2.9e-2}$  
                &   $\pm\num{3.5e-2}$  
                & $\pm\num{4.3e-2}$    
                & $\pm\num{5.2e-2}$    
                \\
\midrule
Magnetic field misalignment 
                                        &     
                                        &    
                                        &     
                                        &     
                                        &     
                                        &   
                                        &   
                                        \\
and scale:    &     
                                        &    
                                        &     
                                        &     
                                        &     
                                        &   
                                        &   
                                        \\

     ~~~~~Bias 
                                        &   $\num{-1.3e-2}$ 
                                        &   $\num{-1.4e-2}$ 
                                        &   $\num{-1.5e-2}$ 
                                        &   $\num{-1.6e-2}$ 
                                        &   $\num{-1.6e-2}$ 
                                        & $\num{-1.7e-2}$ 
                                        & $\num{-1.6e-2}$ 
                                        \\
                                        
~~~~~Uncertainty   
                                        &   $\pm\num{2.0e-4}$ 
                                        &   $\pm\num{2.9e-4}$ 
                                        &   $\pm\num{8.0e-4}$ 
                                        &   $\pm\num{4.8e-4}$ 
                                        &   $\pm\num{5.5e-4}$ 
                                        & $\pm\num{4.8e-4}$ 
                                        & $\pm\num{4.9e-4}$ 
                                        \\

\midrule
Tracker resolution  
                    & $\pm\num{1.6e-3}$ 
                    & $\pm\num{2.1e-3}$ 
                    & $\pm\num{2.8e-3}$ 
                    & $\pm\num{3.8e-3}$ 
                    & $\pm\num{5.3e-3}$ 
                    & $\pm\num{7.0e-3}$ 
                    & $\pm\num{9.5e-3}$
                    \\
\midrule
\midrule
Total systematic uncertainty    
                                &   $\substack{\num{+5.2e-2}\\\num{-4.0e-2}}$ 
                                &   $\substack{\num{+5.7e-2}\\\num{-5.6e-2}}$ 
                                &   $\substack{\num{+5.5e-2}\\\num{-6.2e-2}}$ 
                                &   $\substack{\num{+5.6e-2}\\\num{-5.8e-2}}$ 
                                &   $\substack{\num{+5.5e-2}\\\num{-5.2e-2}}$ 
                                &   $\substack{\num{+11.7e-2}\\\num{-9.5e-2}}$  
                                &   $\substack{\num{+6.9e-2}\\\num{-11.0e-2}}$  
                                \\
\midrule
\midrule

Corrected emittance (mm\,rad) 
                
                &   3.41 
                &   3.66 
                &   3.71 
                &   3.67 
                &   3.71 
                &   3.65 
                &   3.34 
                \\
                
Total uncertainty       
                                                &   $\pm0.06$ 
                        &   $\pm0.07$ 
                        &   $\substack{+0.07\\-0.08}$ 
                        &   $\pm0.08$ 
                        &   $\pm0.09$ 
                        &    $\substack{+0.14\\-0.13}$  
                        &   $\substack{+0.12\\-0.14}$  
                        \\

Total uncertainty  (\%) 
                        &   $\substack{+1.90\\-1.63}$ 
                        &   $\substack{+1.96\\-1.94}$ 
                        &   $\substack{+2.01\\-2.15}$ 
                        &   $\substack{+2.19\\-2.34}$ 
                        &   $\substack{+2.40\\-2.37}$ 
                        & $\substack{+3.97\\-3.49}$   
                        & $\substack{+3.47\\-4.30}$   
                        \\
\bottomrule
\end{tabular}
\end{sidewaystable*}

\begin{figure*}[!ht]
  \begin{center}
    \includegraphics[width=0.8\textwidth]{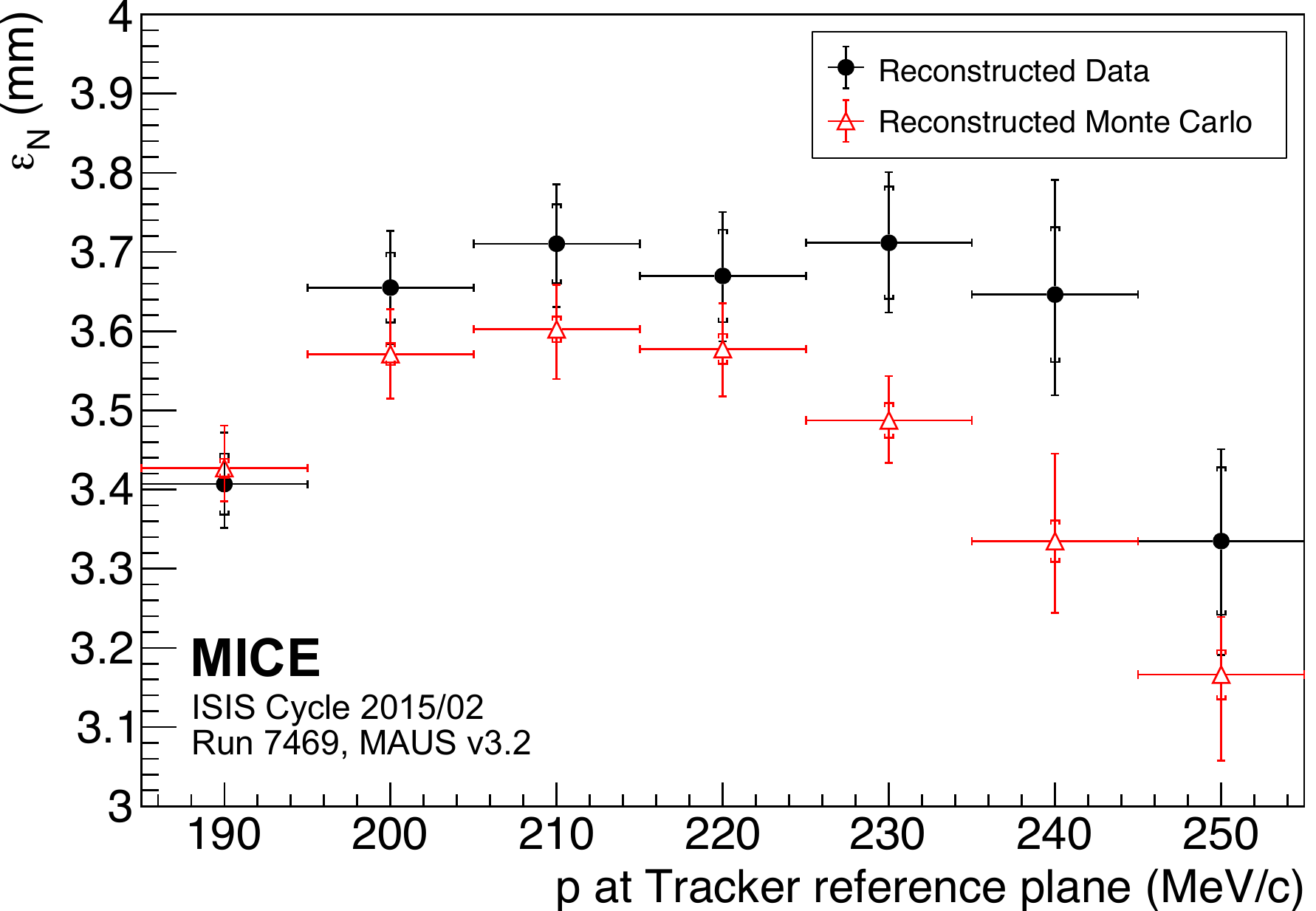}
  \end{center}
  \caption{
    Normalised transverse emittance as a function of total momentum,
    $p$, for data (black, filled circle) and reconstructed Monte Carlo
    (red, open triangle). 
    The inner error bars show the statistical uncertainty.
    The outer error bars show the quadratic sum of the statistical and
    systematic uncertainties.
  } 
  \label{fig:emittance}
\end{figure*}

\subsection{Emittance}
\label{SubSect:Results:Emmttnc}

The normalised transverse emittance as a function of $p$ is shown in
figure~\ref{fig:emittance}. 
The emittance has been corrected for the systematic bias shown in
table~\ref{table-error-summary}. 
The uncertainties plotted are those summarised in
table~\ref{table-error-summary}, where the inner bars represent the
statistical uncertainty and outer bars the total
uncertainty. 
The emittance of the measured muon ensembles (black, filled circle)  is approximately flat in the range $195 \leq p \leq
245$\,MeV/$c$, corresponding to the design momentum of the experiment. 
The mean emittance in this region is $\approx 3.7$\,mm. 
The emittance of the reconstructed Monte Carlo is consistently lower
than that of the data, and therefore gives only an approximate simulation of the beam. 

%% file: 08-Conclusions.tex
\graphicspath{{08-Conclusions/Figures/}}

\section{Conclusions}
\label{Sec:Conclusions}

A first particle-by-particle measurement of the emittance of the MICE Muon Beam was made
using the upstream scintillating-fibre tracking detector in a 4\,T
solenoidal field. 
A total of 24\,660 muons survive the selection criteria. 
The position and momenta of these muons were measured at the reference
plane of the upstream tracking detector. 
The muon sample was divided into 10\,MeV/$c$ bins of total momentum,
$p$, from $185$--$255$\,MeV/$c$ to account for dispersion,
chromaticity, and scraping in apertures upstream of the tracking
detector.
The emittance of the measured muon ensembles is approximately flat from 
$195 \leq p \leq 245$\,MeV/$c$ 
with a mean value of $\approx 3.7$\,mm across this
region.

The total uncertainty on this measurement ranged from
$ \substack{+1.9\\-1.6}$\% to $\substack{+3.5\\-4.3}$\%,
increasing with total momentum, $p$. 
As $p$ increases, the number of muons in the reported ensemble
decreases, increasing the statistical uncertainty. 
At the extremes of the momentum range, a larger proportion of the
input beam distribution is scraped on the aperture of the diffuser. 
This contributes to an increase in systematic uncertainty at the
limits of the reported momentum range. 
The systematic uncertainty introduced by the diffuser aperture
highlights the need to study ensembles where the total momentum, $p$,
is close to the design momentum of the beam line. The total systematic uncertainty on the measured emittance is larger than that on a future measurement of the ratio of emittance before and after an absorber. The measurement is sufficiently precise to demonstrate muon ionization cooling.

The technique presented here represents the first precise
measurement of normalised transverse emittance on a
particle-by-particle basis.  
This technique will be applied to  muon ensembles up- and downstream
of a low-$Z$ absorber, such as liquid hydrogen or lithium hydride, to
measure emittance change across the absorber and thereby to study ionization cooling.

%% file: 11-Acknowledgements.tex
\section*{Acknowledgements}

The work described here was made possible by grants from Department of
Energy and National Science Foundation (USA), the Instituto Nazionale
di Fisica Nucleare (Italy), the Science and Technology Facilities
Council (UK), the European Community under the European Commission
Framework Programme 7 (AIDA project, grant agreement no. 262025, TIARA
project, grant agreement no. 261905, and EuCARD), the Japan Society
for the Promotion of Science, the National Research Foundation of 
Korea (No. NRF-2016R1A5A1013277), and the Swiss National Science
Foundation, in the framework of the SCOPES programme. 
We gratefully acknowledge all sources of support.
We are grateful for the support given to us by the staff of the STFC
Rutherford Appleton and Daresbury Laboratories.
We acknowledge the use of Grid computing resources deployed and
operated by GridPP in the UK, http://www.gridpp.ac.uk/.

This is a post-peer-review, pre-copyedit version of an article published in European Physical Journal C. The final authenticated version is available online at: \url{https://doi.org/10.1140/epjc/s10052-019-6674-y}

\subsubsection*{Data Availability Statement} This manuscript has associated data in a data repository.The data that support the findings of this study are publicly available on the GridPP computing Grid via the data DOIs (the MICE unprocessed data: 10.17633/rd.brunel.3179644; the MICE reconstructed data: 10.17633/rd.brunel.5955850). Publications using the MICE data must contain the following statement: “We gratefully acknowledge the MICE collaboration for allowing us access to their data. Third-party results are not endorsed by the MICE collaboration and the MICE collaboration does not accept responsibility for any errors in the third-party’s understanding of the MICE data.”

%% file: 12-AuthorList.tex
%
\thispagestyle{plain}
\setlength\parindent{0em}

M.~Bogomilov,  R.~Tsenov, G.~Vankova-Kirilova
\\{\it
   Department of Atomic Physics, St.~Kliment Ohridski University of Sofia, Sofia, Bulgaria
}\\

Y.~Song, J.~Tang
\\{\it
Institute of High Energy Physics, Chinese Academy of Sciences, Beijing, China
}\\

Z.~Li
\\{\it
Sichuan University, China
}\\

R.~Bertoni, M.~Bonesini, F.~Chignoli, R.~Mazza
\\{\it
Sezione INFN Milano Bicocca, Dipartimento di Fisica G.~Occhialini, Milano, Italy
}\\

V.~Palladino
\\{\it
Sezione INFN Napoli and Dipartimento di Fisica, Universit\`{a} Federico II, Complesso Universitario di Monte S.~Angelo, Napoli, Italy
}\\

A.~de Bari
\\{\it 
Sezione INFN Pavia and Dipartimento di Fisica, Pavia, Italy
}
% ...
\\

D.~Cline\footnote{Deceased}, X.~Yang
\\{\it
University of California, Los Angeles, CA, USA
}\\

%% file: Emit-Paper-VB.bbl
\providecommand{\href}[2]{#2}\begingroup\raggedright\begin{thebibliography}{10}

\bibitem{Geer:1997iz}
S.~Geer, ``{Neutrino beams from muon storage rings: Characteristics and physics
  potential},'' \href{http://dx.doi.org/10.1103/PhysRevD.57.6989}{{\em Phys.
  Rev.} {\bfseries D57} (1998) 6989--6997},
\href{http://arxiv.org/abs/hep-ph/9712290}{{\ttfamily arXiv:hep-ph/9712290}}.

\bibitem{Apollonio:2002en}
M.~Apollonio {\em et al.}, ``Oscillation physics with a neutrino factory,''
\href{http://arXiv.org/abs/hep-ph/0210192}{{\ttfamily hep-ph/0210192}}.

\bibitem{Neuffer:1994bt}
D.~V. Neuffer and R.~B. Palmer, ``{A High-Energy High-Luminosity $\mu^+ -
  \mu^-$ Collider},''
{\em Conf. Proc.} {\bfseries C940627} (1995) 52--54.

\bibitem{Palmer:2014nza}
R.~B. Palmer, ``{Muon Colliders},''
\href{http://dx.doi.org/10.1142/S1793626814300072}{{\em Rev. Accel. Sci. Tech.}
  {\bfseries 7} (2014) 137--159}.

\bibitem{Boscolo:2018tlu}
M.~Boscolo, M.~Antonelli, O.~R. Blanco-Garcia, S.~Guiducci, S.~Liuzzo,
  P.~Raimondi, and F.~Collamati, ``{Studies of a scheme for Low EMittance Muon
  Accelerator with production from positrons on target},''
\href{http://arxiv.org/abs/1803.06696}{{\ttfamily arXiv:1803.06696
  [physics.acc-ph]}}.

\bibitem{2012acph.book.....L}
S.~Y. {Lee}, \href{http://dx.doi.org/10.1142/8335}{{\em {Accelerator Physics
  (Third Edition)}}}.
\newblock World Scientific Publishing Co, 2012.

\bibitem{PhysRevLett.64.2901}
S.~Schr\"oder, R.~Klein, N.~Boos, M.~Gerhard, R.~Grieser, G.~Huber,
  A.~Karafillidis, M.~Krieg, N.~Schmidt, T.~K\"uhl, R.~Neumann, V.~Balykin,
  M.~Grieser, D.~Habs, E.~Jaeschke, D.~Kr\"amer, M.~Kristensen, M.~Music,
  W.~Petrich, D.~Schwalm, P.~Sigray, M.~Steck, B.~Wanner, and A.~Wolf, ``First
  laser cooling of relativistic ions in a storage ring,''
  \href{http://dx.doi.org/10.1103/PhysRevLett.64.2901}{{\em Phys. Rev. Lett.}
  {\bfseries 64} (Jun, 1990) 2901--2904}.
  \url{http://link.aps.org/doi/10.1103/PhysRevLett.64.2901}.

\bibitem{PhysRevLett.67.1238}
J.~S. Hangst, M.~Kristensen, J.~S. Nielsen, O.~Poulsen, J.~P. Schiffer, and
  P.~Shi, ``Laser cooling of a stored ion beam to 1 m{K},''
  \href{http://dx.doi.org/10.1103/PhysRevLett.67.1238}{{\em Phys. Rev. Lett.}
  {\bfseries 67} (Sep, 1991) 1238--1241}.
  \url{http://link.aps.org/doi/10.1103/PhysRevLett.67.1238}.

\bibitem{doi:10.1063/1.329218}
P.~J. Channell, ``Laser cooling of heavy ion beams,''
  \href{http://dx.doi.org/10.1063/1.329218}{{\em Journal of Applied Physics}
  {\bfseries 52} no.~6, (1981) 3791--3793},
  \href{http://arxiv.org/abs/http://dx.doi.org/10.1063/1.329218}{{\ttfamily
  http://dx.doi.org/10.1063/1.329218}}.
  \url{http://dx.doi.org/10.1063/1.329218}.

\bibitem{Marriner:2003mn}
J.~Marriner, ``{Stochastic cooling overview},''
  \href{http://dx.doi.org/10.1016/j.nima.2004.06.025}{{\em Nucl. Instrum.
  Meth.} {\bfseries A532} (2004) 11--18},
\href{http://arxiv.org/abs/physics/0308044}{{\ttfamily arXiv:physics/0308044
  [physics]}}.

\bibitem{1063-7869-43-5-R01}
V.~V. Parkhomchuk and A.~N. Skrinsky, ``Electron cooling: 35 years of
  development,'' {\em Physics-Uspekhi} {\bfseries 43} no.~5, (2000) 433--452.
  \url{http://stacks.iop.org/1063-7869/43/i=5/a=R01}.

\bibitem{cooling_methods}
A.~N. Skrinsky and V.~V. Parkhomchuk, ``{Cooling Methods for Beams of Charged
  Particles. (In Russian)},'' {\em Sov. J. Part. Nucl.} {\bfseries 12} (1981)
  223--247.
[Fiz. Elem. Chast. Atom. Yadra12,557(1981)].

\bibitem{Neuffer:1983xya}
D.~Neuffer, ``{Principles and Applications of Muon Cooling},''
{\em Conf.Proc.} {\bfseries C830811} (1983) 481.

\bibitem{Neuffer:1983jr}
D.~Neuffer, ``{Principles and Applications of Muon Cooling},''
{\em Part. Accel.} {\bfseries 14} (1983) 75--90.

\bibitem{MICE-WWW}
{The MICE collaboration}, ``{INTERNATIONAL MUON IONIZATION COOLING
  EXPERIMENT}.'' \url{http://mice.iit.edu}.

\bibitem{Apollonio:2008aa}
{\bfseries ISS Accelerator Working Group} Collaboration, M.~Apollonio {\em et
  al.}, ``{Accelerator design concept for future neutrino facilities},''
  \href{http://dx.doi.org/10.1088/1748-0221/4/07/P07001}{{\em JINST} {\bfseries
  4} (2009) P07001},
\href{http://arxiv.org/abs/0802.4023}{{\ttfamily arXiv:0802.4023
  [physics.acc-ph]}}.

\bibitem{Rosenzweig:BeamPhysics}
J.~B. Rosenzweig, {\em Fundamentals of Beam Physics}.
\newblock Oxford University Press, 2003.

\bibitem{Booth:2012qz}
C.~N. Booth {\em et al.}, ``{The design, construction and performance of the
  MICE target},'' \href{http://dx.doi.org/10.1088/1748-0221/8/03/P03006}{{\em
  JINST} {\bfseries 8} (2013) P03006},
\href{http://arxiv.org/abs/1211.6343}{{\ttfamily arXiv:1211.6343
  [physics.ins-det]}}.

\bibitem{Booth:2016lno}
C.~N. Booth {\em et al.}, ``{The design and performance of an improved target
  for MICE},'' \href{http://dx.doi.org/10.1088/1748-0221/11/05/P05006}{{\em
  JINST} {\bfseries 11} no.~05, (2016) P05006},
\href{http://arxiv.org/abs/1603.07143}{{\ttfamily arXiv:1603.07143
  [physics.ins-det]}}.

\bibitem{Bogomilov:2012sr}
{\bfseries MICE} Collaboration, M.~Bogomilov {\em et al.}, ``{The MICE Muon
  Beam on ISIS and the beam-line instrumentation of the Muon Ionization Cooling
  Experiment},'' \href{http://dx.doi.org/10.1088/1748-0221/7/05/P05009}{{\em
  JINST} {\bfseries 7} (2012) P05009},
\href{http://arxiv.org/abs/1203.4089}{{\ttfamily arXiv:1203.4089
  [physics.acc-ph]}}.

\bibitem{Adams:2013lba}
{\bfseries MICE} Collaboration, D.~Adams {\em et al.}, ``{Characterisation of
  the muon beams for the Muon Ionisation Cooling Experiment},''
  \href{http://dx.doi.org/10.1140/epjc/s10052-013-2582-8}{{\em Eur. Phys. J.}
  {\bfseries C73} no.~10, (2013) 2582},
\href{http://arxiv.org/abs/1306.1509}{{\ttfamily arXiv:1306.1509
  [physics.acc-ph]}}.

\bibitem{Adams:2015wxp}
{\bfseries MICE} Collaboration, M.~Bogomilov {\em et al.}, ``{Pion
  contamination in the MICE muon beam},''
  \href{http://dx.doi.org/10.1088/1748-0221/11/03/P03001}{{\em JINST}
  {\bfseries 11} no.~03, (2016) P03001},
\href{http://arxiv.org/abs/1511.00556}{{\ttfamily arXiv:1511.00556
  [physics.ins-det]}}.

\bibitem{Bertoni:2010by}
{\bfseries MICE} Collaboration, R.~Bertoni {\em et al.}, ``{The design and
  commissioning of the MICE upstream time-of-flight system},''
  \href{http://dx.doi.org/10.1016/j.nima.2009.12.065}{{\em Nucl.Instrum.Meth.}
  {\bfseries A615} (2010) 14--26},
\href{http://arxiv.org/abs/1001.4426}{{\ttfamily arXiv:1001.4426
  [physics.ins-det]}}.

\bibitem{MICE:Note:286:2010}
R.~Bertoni, M.~Bonesini, A.~de~Bari, G.~Cecchet, Y.~Karadzhov, and R.~Mazza,
  ``{The construction of the MICE TOF2 detector}.'' \url{
  http://mice.iit.edu/micenotes/public/pdf/MICE0286/MICE0286.pdf }, 2010.

\bibitem{Cremaldi:2009zj}
L.~Cremaldi, D.~A. Sanders, P.~Sonnek, D.~J. Summers, and J.~Reidy, Jr, ``{A
  Cherenkov Radiation Detector with High Density Aerogels},''
  \href{http://dx.doi.org/10.1109/TNS.2009.2021266}{{\em IEEE Trans. Nucl.
  Sci.} {\bfseries 56} (2009) 1475--1478},
\href{http://arxiv.org/abs/0905.3411}{{\ttfamily arXiv:0905.3411
  [physics.ins-det]}}.

\bibitem{Ellis:2010bb}
M.~Ellis {\em et al.}, ``{The design, construction and performance of the MICE
  scintillating fibre trackers},''
  \href{http://dx.doi.org/10.1016/j.nima.2011.04.041}{{\em Nucl. Instrum.
  Meth.} {\bfseries A659} (2011) 136--153},
\href{http://arxiv.org/abs/1005.3491}{{\ttfamily arXiv:1005.3491
  [physics.ins-det]}}.

\bibitem{Ambrosino2009239}
F.~Ambrosino {\em et al.}, ``{Calibration and performances of the KLOE
  calorimeter},''
\href{http://dx.doi.org/10.1016/j.nima.2008.08.097}{{\em Nucl. Instrum. Meth.}
  {\bfseries A598} (2009) 239--243}.

\bibitem{Asfandiyarov:2016erh}
R.~Asfandiyarov {\em et al.}, ``{The design and construction of the MICE
  Electron-Muon Ranger},''
  \href{http://dx.doi.org/10.1088/1748-0221/11/10/T10007}{{\em JINST}
  {\bfseries 11} no.~10, (2016) T10007},
\href{http://arxiv.org/abs/1607.04955}{{\ttfamily arXiv:1607.04955
  [physics.ins-det]}}.

\bibitem{Adams:2015eva}
{\bfseries MICE} Collaboration, D.~Adams {\em et al.}, ``{Electron-Muon Ranger:
  performance in the MICE Muon Beam},''
  \href{http://dx.doi.org/10.1088/1748-0221/10/12/P12012}{{\em JINST}
  {\bfseries 10} no.~12, (2015) P12012},
\href{http://arxiv.org/abs/1510.08306}{{\ttfamily arXiv:1510.08306
  [physics.ins-det]}}.

\bibitem{Dobbs:2016ejn}
A.~Dobbs, C.~Hunt, K.~Long, E.~Santos, M.~A. Uchida, P.~Kyberd, C.~Heidt,
  S.~Blot, and E.~Overton, ``{The reconstruction software for the MICE
  scintillating fibre trackers},''
  \href{http://dx.doi.org/10.1088/1748-0221/11/12/T12001}{{\em JINST}
  {\bfseries 11} no.~12, (2016) T12001},
\href{http://arxiv.org/abs/1610.05161}{{\ttfamily arXiv:1610.05161
  [physics.ins-det]}}.

\bibitem{Blot:2011zz}
S.~Blot, ``{Proton Contamination Studies in the MICE Muon Beam Line},'' {\em
  {Proceedings 2nd International Particle Accelerator Conference (IPAC 11) 4-9
  September 2011, San Sebastian, Spain}} (2011) .

\bibitem{miceDataArchive}
{The MICE Collaboration}, ``{The MICE RAW Data}.''
  \url{doi:10.17633/rd.brunel.3179644}.
\newblock MICE/Step4/07000/07469.tar.

\bibitem{G4BeamLine:WWW}
T.~Roberts {\em et al.}, ``{G4beamline; a ``Swiss Army Knife'' for Geant4,
  optimized for simulating beamlines}.'' \url{
  http://public.muonsinc.com/Projects/G4beamline.aspx }.

\bibitem{MICE:Note:439:2014}
D.~Rajaram and C.~Rogers, ``The mice offline computing capabilities.'' \url{
  http://mice.iit.edu/micenotes/public/pdf/MICE0439/MICE0439.pdf }, 2014.
\newblock MICE Note 439.

\bibitem{Ago03}
{\bfseries GEANT4} Collaboration, S.~Agostinelli {\em et al.}, ``Geant4: A
  simulation toolkit,''
{\em Nuclear Instruments and Methods in Physics Research A} {\bfseries 506}
  (2003) 250--303.

\bibitem{Allison:2006ve}
J.~Allison {\em et al.}, ``{Geant4 developments and applications},''
\href{http://dx.doi.org/10.1109/TNS.2006.869826}{{\em IEEE Trans. Nucl. Sci.}
  {\bfseries 53} (2006) 270--278}.

\bibitem{Brun:1997pa}
R.~Brun and F.~Rademakers, ``{ROOT: An object oriented data analysis
  framework},''
\href{http://dx.doi.org/10.1016/S0168-9002(97)00048-X}{{\em Nucl. Instrum.
  Meth.} {\bfseries A389} (1997) 81--86}.

\bibitem{MICE:Note:268:2009}
J.~Cobb, ``{Statistical Errors on Emittance Measurements}.'' \url{
  http://mice.iit.edu/micenotes/public/pdf/MICE341/MICE268.pdf }, 2009.

\bibitem{MICE:Note:341:2011}
J.~Cobb, ``{Statistical Errors on Emittance and Optical Functions}.'' \url{
  http://mice.iit.edu/micenotes/public/pdf/MICE341/MICE341.pdf }, 2011.

\bibitem{Unpublished:MICE:Note:2015}
J.~H. Cobb, ``Statistical errors on emittance.'' Private communication, 2015.

\bibitem{Hunt-note-in-progress}
C.~Hunt, ``Private communication,''.

\bibitem{comsol}
{\url{HTTP://WWW.COMSOL.COM/}}.

\bibitem{Lyons:0305-4470-25-7-035}
L.~Lyons, ``On estimating systematic errors from repeated measurements,'' {\em
  Journal of Physics A: Mathematical and General} {\bfseries 25} no.~7, (1992)
  1967. \url{http://stacks.iop.org/0305-4470/25/i=7/a=035}.

\end{thebibliography}\endgroup
